\documentclass[aps,preprint,showpacs,showkeys]{revtex4}%
\usepackage{amsmath,bm}
\usepackage{amsmath}
\usepackage{amsfonts}
\usepackage{amssymb}
\usepackage{graphicx}%
\begin{document}
\date{\today }
\title{ Conformal symmetry and light flavor baryon spectra       }
\author{M.\ Kirchbach, C.\ B.\ Compean}
\affiliation{$^2$Instituto de F\'{\i}sica, \\
         Universidad Aut\'onoma de San Luis Potos\'{\i},\\
         Av. Manuel Nava 6, San Luis Potos\'{\i}, S.L.P. 78290, M\'exico}

\newcommand{\csch}{\textrm{ csch }}
\newcommand{\sech}{\textrm{ sech }}
\newcommand{\arccot}{\textrm{ arccot }}
\newcommand{\arccoth}{\textrm{ arccoth }}
\newcommand{\e}{\textrm { e}}

\newcommand{\be}{\begin{eqnarray}}
\newcommand{\ee}{\end{eqnarray}}
\newcommand{\nn}{\nonumber}






\begin{abstract}
The degeneracy among parity pairs systematically observed
in the $N$ and $\Delta$ spectra is interpreted to hint on a
possible conformal symmetry realization in the light flavor baryon 
sector in line with AdS$_5$/CFT$_4$. 
The case is made by showing that  all the observed
$N$ and $\Delta$ resonances with masses below 2500 MeV 
distribute fairly well each over the first levels of a
unitary representation of the conformal group, a representation that
covers the spectrum of a quark-diquark system, placed directly  
on a conformally compactified Minkowski spacetime,
{\bf R}$^1\otimes S^3$, as approached from the the  AdS$_5$ cone.
The free geodesic motion on the  $S^3$ 
manifold is described by means of the 
scalar conformal equation there, which is of the Klein-Gordon type. 
The equation is then gauged by  
the ``curved'' Coulomb  potential that  has the form of a 
cotangent function. Conformal symmetry is not exact,
this because the gauge potential slightly modifies
the  conformal centrifugal barrier of the free geodesic motion.
Thanks to this, the degeneracy between $P_{11}-S_{11}$ pairs 
from same level is relaxed, while the remaining states belonging to same level
remain practically degenerate.
The model describes the correct mass ordering in the
$P_{11}-S_{11}$ pairs through the nucleon spectrum as a combined  
effect of the above conformal symmetry breaking, on the one side,
and a parity change of the diquark from a scalar at low masses,
to a pseudoscalar at higher masses, on the other.
The quality of the wave functions is illustrated by 
calculations of realistic mean-square charge radii 
and electric charge form-factors on the examples of the
proton, and the protonic $P_{11}(1440)$, and $S_{11}(1535)$ resonances.
The scheme also allows for a prediction of
the dressing function of an effective 
instantaneous gluon propagator from the Fourier transform of 
the gauge potential. We find a dressing function that is finite in the 
infrared and tends to zero at infinity.
\end{abstract}

\pacs{12.39.Jh, 24.85.+p}

\keywords{AdS/CFT, higher spins, electric mean square charge radii}
\maketitle

\section{Introduction}
Understanding the systematics of  high-spin states is among the serious 
challenges in quark spectroscopy \cite{Lee},\cite{Afonin}. 
This because the number of resonances
predicted by the traditional quark models
based upon the full Hilbert space of six spin-flavor degrees of freedom
and the lowest symmetry of the radial wave functions, the rotational 
invariance,
\cite{Capstick}  significantly exceeds the number of the states observed 
so far \cite{PART}. 
The resonance deficit,  termed to as ``missing'' states, 
is still awaiting for explanation. Quark-diquark (q--(qq)) 
models \cite{Wilczek} based on a diquark with limited angular momentum
values, carry 
reduced spin-flavor degrees of freedom, and are obvious 
candidates for providing a lesser
number of ``missing'' states, an option taken into consideration by
several authors \cite{Elena},~\cite{tnk}. 
Additional restrictions on the quantum numbers
of the $q$-$(qq)$ excitations can come from 
imposing on the spatial wave functions a symmetry higher than SO(3)$_L$.
A natural candidate would be the,
admittedly approximate, global conformal symmetry
of the QCD Lagrangian in the light flavor sector. 
It is the goal of the present work to examine consequences of conformal 
symmetry for the systematics
of the $N$ and $\Delta$ spectra in
constructing radial wave functions of a $q$--$(qq)$
system  in accord with that very  symmetry. The aim is to pick up
from the full Hilbert space isospin by isospin 
those $N$ and $\Delta$ resonances  
whose quantum numbers would fit
into an irreducible representation
of the conformal group (termed to as conformal band). 
The observation of these states
can then serve as a signature for conformal symmetry
realization in the infrared.
Below we design such a model in reference to the AdS$_5$/CFT$_4$ concept.
We shall see that the model allows to interpret the
systematically observed degeneracy among parity pairs 
in the $N$ and $\Delta$ spectra as a hint on a possible 
realization of conformal symmetry in light flavor baryon spectra. 
\noindent
Implementations of conformal symmetry and
the AdS/CFT concept to hadron physics have been 
pioneered in refs.~\cite{Br1},\cite{Br2}  within the framework of  light-front 
QCD. Further interesting studies can be found, 
among others, in refs.~\cite{dong}, \cite{Park},
\cite{Nawa}, \cite{hashimoto}.  Specifically in refs.~\cite{Br1},\cite{Br2},
a conformally invariant  Hamiltonian
has been successfully employed in the construction of 
spatial wave functions for both mesons and baryons.
We here implement conformal symmetry into a quark Hamiltonian
in a position space of a finite volume.
This is achieved in placing the $q$-$(qq)$ 
system directly on the AdS$_5$ boundary, which is the AdS$_5$ cone,
considered as conformally compactified to 
$S^1\otimes S^3$ \cite{Nicolai}, or, {\bf R}$^{1}\otimes S^{3} $, at a
microscopic scale  \cite{Mack}. 
According to \cite{Mack}, correlation functions of 
CFT$_4$ on regular Minkowski spacetime, ${\mathcal M}$={\bf R}$^{1+3}$,
 can be analytically continued to the 
full Einstein universe, this because {\bf R}$^{1+3}$ 
can be conformally mapped on {\bf R}$^1\otimes S^3$.
The implication of this important observation is that
each state of the CFT on 
{\bf R}$^1\otimes S^3$ can be brought into unique correspondence with a 
state of the brane theory on $AdS_{5}\otimes S^5$. 
Consequences  on thermal states have been worked out in ref.~\cite{Horow}. 
We here examine consequences for the systematics
of the $N$ and $\Delta$ spectra.

\noindent
The paper is organized as follows.
In the next section we highlight the procedure of
conformal compactification of Minkowski spacetime
along the line of ref.~\cite{Gibbons} and present the
conformal equation which will be applied to
a  quark-diquark (q-qq) model of light baryons in section 3.
There, we calculate the $N$ and $\Delta$ 
spectra, the mean square charge radii and 
the corresponding electric charge form-factors
of some of the excited states.
Section 4 is devoted to the  design of a dressing function of 
an (instantaneous) effective gluon 
propagator as a Fourier transform of the gauge potential on
$S^3$.
The paper closes with brief conclusions.

\section{ Conformal compactification of the AdS$_5$ cone
to $S^1\otimes S^3\sim${\bf R}$^1\otimes S^3$, wave equations,
and cotangent confinement potential} 

Applications of brane theory to hadron physics have
acquired considerable attention in recent times. Such a
possibility arose in effect of the intriguing observation
\cite{Nicolai}-\cite{Mack} that the asymptotic horizon geometry
of the Dirichlet three-brane (D3) of the
IIB superstring in ten dimensions considered  on 
a  $ AdS_{5}\otimes S^{5 }$ background,
admits a superalgebra that is identical to the super-conformal algebra
of the corresponding four dimensional world-volume 
field theory  when gravity is decoupled.  
The D3 brane theory has $SU(2,2/4)$  as underlying
superconformal symmetry  whose bosonic isometry 
 $SU(2,2)\otimes SU(4)$, is  locally isomorphic to
$ SO(2,4)\otimes SO(6)$ \cite{Frr_Frnsdl}.
This  group happens to coincide with
the isometry group of the corresponding 
asymptotic horizon background, $AdS_5\otimes S^5$, 
a mathematical coincidence
that was suggestive of a duality between supergravity 
around the horizon background, on the one side,
and superconformal brane dynamics, on the other.
In such a  type of strong-weak  duality, one expects the 
fundamental supergraviton degrees of freedom  
to show up as bound states in the 
non-perturbative regime of the corresponding world-volume theory. 
As a reminder,  D3-branes  solve the ten-dimensional
supergravity equations of motion and 
in having a $(1+3)$ dimensional world volume,
are surrounded by six  transverse dimensions. 
In polar coordinates five of the $SO(6)$ dimensions are accounted for
in terms of the angular coordinates parameterizing the $S^5$ 
hypersphere, while the sixth hyper radial coordinate is to 
become holographic with respect to $AdS_{5}$ in the near 
horizon geometry. There, the branes low-energy
effective description, the Yang-Mills theory, becomes 
equivalent to type II B string, a reason for which
a  duality between   $AdS_{5}$ and 4d 
supersymmetric  Yang-Mills theory has been conjectured by Maldacena
 ~\cite{Wilson}.
Maldacena's conjecture \cite{Wilson}  
woke up expectations that zero temperature super Yang-Mills theory 
residing in the conformal $AdS_{5}$ boundary is likely
to capture some of the essential  features of high-temperature 
three-dimensional QCD \cite{Seiberg}--\cite{vijay}.  
Within this context, testing AdS/CFT reduces to
the calculation of observables within that very framework 
and their comparison to the corresponding 
Lattice results in 3d  QCD.
The duality
between D3-brane bulk supergravity and super Yang-Mills open string theories
on the conformal boundary  
of $AdS_{5}$ space-time  implies  that  each CFT state, 
(among them the QCD states in the light flavor sector)
can be put in correspondence  to a state in the 
supergravity approximation to string theory
on $AdS_{5}\otimes S^5$ meaning that spectra 
in both theories should come out same.
As long as the isometry group of the conformal  
$AdS_{5}$  boundary  is the conformal group
$SO(2,4)$,  the requirement is that the  quantum states in the
theory on that  boundary  should populate $SO(2,4)$ 
unitary representations \cite{Seiberg}.
This is an essential restriction which 
strongly limits the number of  theories in line with the AdS/CFT 
and is suggestive of the construction of quark models 
that respect global conformal invariance.
Conformal symmetry is independently 
to a good approximation a global symmetry of 
the QCD Lagrangian in the light-flavor sector, a reason 
for which one can expect  spectroscopic data 
on the light flavor baryons, 
the nucleon and the $\Delta$, to be especially
appropriate in examining the AdS/CFT concept.
The implementation of conformal symmetry by a quark model 
has to combine with the confinement phenomenon, whose  
description is one of the
major goals in the physics of hadrons. Confinement implies
exclusion of 
scattering states and favoring  discrete bound states alone.   
Putting systems on finite volumes
is a standard strategy of spectrum discretization \cite{barbon}.
There is a variety of geometries appropriate for  
preserving  the conformal symmetry by the resulting Hamiltonians,
the three dimensional sphere, $S^3$, being one of them 
\cite{Mack},  \cite{barbon}, \cite{Gibbons}.
It has been shown that a geometry containing $S^3$, such as the 
 $S^1\otimes S^3$  manifold,
can be  approached departing
directly from AdS$_5$ \cite{Nicolai}, \cite{Gibbons1}, \cite{Gibbons}.
The five dimensional manifold AdS$_5$ is  defined as a 
{\bf R}$^{2+4}$ subspace according to 
\begin{equation}
u^2+v^2 -x_1^2-x_2^2-x_3^2-x_4^2=\rho^2,
\end{equation}
where $\rho $ is a fixed parameter.  The boundary at infinity of this space
is identified with the $AdS_5$ cone
\begin{equation}
u^2+v^2 -x_1^2-x_2^2-x_3^2-x_4^2=0.
\end{equation}
Flat four-dimensional (4d) Minkowski space time can then be thought of as the 
intersection of the null hyperplane $v^2-x_4^2=0$ with the
$AdS_5$ cone, 
\begin{equation}
v^2-x_4^2=u^2- x_1^2-x_2^2-x_3^2=0, 
\label{light_rays}
\end{equation}
in which case $u$ and ${\vec x}=$column$(x_1,x_2,x_3)$
in turn assume the r\'oles of time and position vector in
{\bf R}$^{1+3}$, respectively. A quark system placed on such a 5d cone
can then be described by means of the light-front formalism. This path
has been taken by refs.~\cite{Br1},~\cite{Br2} and culminated to the
holographic light-front QCD.\\

\noindent 
The path taken by the present investigation is rather to
consider Minkowski space as emerging from an
AdS$_5$ cone alternatively  patterned after,
\begin{equation}
u^2+v^2=x_1^2+x_2^2+x_3^2+x_4^2=R^2, \quad R\not=0, 
\label{null_rays}
\end{equation}
in which case it  compactifies to 
${\mathcal M}^{\ast( 1+3)}=S^1\otimes S^3$.
The 
$S^1\otimes S^3$ manifold then describes  
the particular set of {\bf R}$^{2+4}$ null rays
associated with  eq.~(\ref{null_rays}).
As long as the isometry group,
$SO(2)\otimes SO(4)$,  of $S^1\otimes S^3$,
is a subgroup of the conformal group $SO(2,4)$ of 
regular (1+3) Minkowski space, ${\mathcal M}$, i.e.
$SO(2)\otimes SO(4)\subset SO(2,4)$, 
eq.~(\ref{null_rays}) is also referred to as
conformal compactification  of Minkowski space time. 
Now one can parametrize the $S^1\otimes S^3$ manifold by the four angles 
$\tau, \chi, \theta, \varphi$ in accordance with
\begin{eqnarray}
u+iv=Re^{i\tau }\, &\quad& x_1+ix_2=R\sin\chi \sin\theta e^{i\varphi},\quad
x_3=R\sin \chi \cos \theta, \nonumber\\
x_4^2+{\mathbf r}^2=R^2,&\quad&
 r= |{\mathbf r}|=R\sin\chi,\quad \chi =\sin^{-1}r\sqrt{\kappa},
\quad \kappa=\frac{1}{R^2},
\label{RW_prmtrz}
\end{eqnarray}
where  $R$  is the $S^3$ hyper-radius, and $\kappa $ the curvature. 
According to ref.~\cite{Gibbons}, this map takes at a microscopic scale the
flat space Minkowski metric to the metric of
Einstein's  {\bf R}$^1\otimes S^3$ cylinder, 
\begin{equation}
ds^2=\Omega ^{-2}
(-{\mathrm d}\tau^2 +
{\mathrm d}\chi^2 +\sin^2\chi\left(
{\mathrm  d}\theta^2 +\sin^2\theta 
{\mathrm d} \varphi ^2)\right), 
\label{metric}
\end{equation}
with $\Omega$ being the conformal factor. In this way, one establishes
the relationship,
$S^1\otimes S^3\simeq ${\bf R}$^{1}\otimes S^3$.

\subsection{Free geodesic motion on $S^3$ and the conformal free rigid rotor.}

Within the metric of eq.~(\ref{metric}), 
and the conformal factor being absorbed by the 
wave functions,  the following
conformally invariant massless scalar field equation 
has been found in ref.~\cite{Gibbons},\cite{Blaschke}
\begin{equation}
-\hbar^2 {\widehat \Box }\psi +\mu^2 \psi =0.
\label{conf_eq}
\end{equation}
Here, $\mu^2$ is the conformal constant (Ricci scalar) \cite{Gibbons}, 
${\widehat \Box}$  stands for the angular part of the 
4d Laplace-Beltrami operator, which we here choose to express in terms of
${\mathcal K}^2$, and  ${\mathbf L}^2$,    
the operators of the squared
four- and three dimensional
angular momenta as,
 \begin{eqnarray}
{\widehat \Box} &=&  
-\frac{1 }{R^2}\frac{\partial ^2}{\partial \tau ^2}-
\frac{1}{R^2}{\mathcal K}^2,\nonumber\\
-{\mathcal K}^2&=&  \left[\frac{1}{\sin^2\chi }
\frac{\partial }{\partial \chi}
\sin^2\chi \frac{\partial }{\partial \chi } -
\frac{{\mathbf  L}^2 (\theta ,\varphi ) }{\sin^2 \chi }\right].
\label{lpls_5}
\end{eqnarray}
Furthermore, $\chi \in [0,\pi ]$ is the second polar angle on $S^3$.
The ${\mathcal K}^2$ eigenstates, $|Klm>$, are well known to
belong to irreducible  $SO(4)$ representations of the type  
$\left(\frac{K}{2},\frac{K}{2} \right)$, and the quantum numbers, 
$K$, $l$, and $m$ define the eigenvalues of the respective  
four--, three-- and two--dimensional angular momentum operators
upon the  states \cite{Kim_Noz}. The ${\mathcal K}^2$ eigenvalues, 
$\lambda _K$, upon $|Klm>$, are known too and given by, 
\begin{eqnarray}
{\mathcal K}^2 \vert K l m \rangle = {\lambda }_K
\vert K l m \rangle,
&& {\lambda  }_K=K(K+2), \quad
\vert Klm\rangle \in \left( \frac{K}{2},\frac{K}{2} \right),
\quad  \quad K\in [0, \infty),
\nonumber\\
{\mathbf L}^2 |Klm\rangle =l(l+1)|Klm\rangle ,
&\quad&  L_z|Klm>=m|Klm\rangle,\nonumber\\
l\in [0,K], &\quad&  m\in [-l,l].
\label{Casimir_O4}
\end{eqnarray}

\begin{quote}
The infinite series of  solutions of 
eq.~(\ref{Casimir_O4}) constitute 
an $\infty$d unitary representation of the conformal group that has been
built up from the
eigenstates of its little group, $SO(4)_{K}$ \cite{Frr_Frnsdl},
\cite{Wybourne}. This irrep will be frequently 
termed to as a ``conformal band''.
Therefore, the conformal symmetry aspect of AdS/CFT   
is adequately captured by the ${\mathcal K}^2$ eigenvalue 
problem, which in this fashion qualifies  
as a suitable departure point toward the description of  
conformal excitation modes in two-body systems.
\end{quote}

Independently, conformal symmetry is also to a good approximation global
symmetry of the QCD Lagrangian in the light flavor sector, one more reason
why employing the conformally invariant AdS/CFT scenario from above
in modeling $N$ and $\Delta $ excitations should be of interest.
Upon factorizing the $\tau $ dependence of the solution to
eq.~(\ref{conf_eq}) as 
$\exp {\left( i\frac{
{\mathcal E}}
{\hbar\sqrt{\kappa } } \tau \right)
}$,  one arrives at the following
angular equation  on $S^3$,
\begin{eqnarray}
{\Big[} \kappa {\hbar^2}{\mathcal K}^2 
&-& {\mathcal E}^2+\mu^2  {\Big]} 
{\mathcal S}  (\chi)=0.
\label{chi_eq}
\end{eqnarray}
Though in angular variables, algebraically the conformal 
eq.~(\ref{conf_eq}),
has the form of  a Klein-Gordon equation which provides a 
relativistic description of the free geodesic motion on $S^3$
in terms of  the eigenvalue problem of the squared four dimensional 
angular momentum.
The spectrum of eq.~(\ref{chi_eq}) reads,
\begin{equation}
{\mathcal E}^2_K -\mu^2 =\kappa {\hbar^2}K\left(K+2\right).
\label{4D_rotator}
\end{equation}
and represents no more but what one could term to as   
the spectrum of the ``conformal free rigid  rotor''.
On the unit sphere, $\kappa =1$,
the solutions contain the 
Gegenbauer polynomials. The normalized total angular functions for this case, 
i.e. the ${\mathcal K}^2$ eigenstates in 
eqs.~(\ref{Casimir_O4}),~(\ref{chi_eq})
are then given by,
\begin{equation}
|Klm>={\mathcal S}(\chi)Y_l^m(\theta, \varphi)\longrightarrow
N_{Klm}\sin^l\chi C_{K-l}^{l+1}(\cos \chi )Y_l^m(\theta, \varphi)\equiv 
Y_{Klm}(\chi, \theta, \varphi)
\label{K_qnbr}
\end{equation}
where $Y_{Klm}(\chi, \theta, \varphi)$ are  well  known 
4d  hyper-spherical harmonics.

As a more specialized spectroscopic reading to eqs.~({\ref{chi_eq}), 
one can say that the ${\mathcal K}^2$ eigenvalue problem
has $SO(4)$ as potential algebra, and 
the conformal group as dynamical symmetry. \\

\noindent
Had one started instead with AdS$_4$, one would have ended up with the
spectrum of the relativistic 3d free rigid   rotor,
$E^2-\mu^2 = \hbar^2\kappa l(l+1)$. Such a spectrum  may be associated with
Regge trajectories \cite{Bohm}, a reason for the frequently
preferred plots of resonance excitations on a 
mass$^2$/$l$ grid beyond the strict S-matrix concept of Regge trajectories
\cite{dong}, \cite{Anis}. \\

\noindent
In the next section we shall  introduce an interaction
in eq.~(\ref{chi_eq}).

\subsection{The conformal interacting rigid rotor.} 
The next step is introducing  interaction on $S^3$ in such a manner as
to respect the conformal symmetry of the spectrum of the free geodesic motion.
From potential theory it is known that such an interaction has to satisfy
the Laplace-Beltrami equation on the manifold under consideration,
meaning that it has to be a harmonic function there \cite{Axler}. 
Harmonic functions are known for  their property to respect the symmetry
of the Laplacian which is the conformal symmetry of 
the respective $d$-dimensional space. Specifically on $S^3$, 
it has been known for a long time \cite{Schr40}, \cite{Higgs} that $\cot\chi $,
occasionally termed to as ``curved'' Coulomb potential \cite{Barut},
is a harmonic function and suitable as a conformal potential. 
According to the $S^3$ parametrization in eq.~(\ref{RW_prmtrz})
the potential under consideration is of finite range,
\begin{equation}
\cot\chi =\frac{x_4}{r}, \quad r\in [0,R], \quad x_4\in [-R,R],
\end{equation}
and describes interactions on $S^3$ in their projection
onto the equatorial plane, a 3d position space of a finite volume.
In due course we shall reveal importance of the finite range 
of the confinement potential on various spectroscopic observables
in the baryon sector. 
 The conformal character of
the cotangent on $S^3$ is independently illustrated by the fact that
it also describes an exact string solution corresponding to
a D3 brane with transverse dimensions conformally wrapped over $S^3$,
a result due to refs.~\cite{Hrwtz}, \cite{Papad_Tseytlin}.
There, a broad class of
exact string solutions have been constructed
by wrapping transversal dimensions of
fundamental strings over curved spaces and solving the 
corresponding curved space Laplace-Beltrami equations for harmonic functions.
Specifically on $S^3$, the harmonic function 
$K(\chi)$ (in the notations of ref.~\cite{Hrwtz})
obtained as a solution of the 4d Laplace-Beltrami equation,
\begin{equation}
{\widehat \Box} K(\chi) = 0, \quad K(\chi )= {\bar a}+m\cot\chi .
\label{Horus2}
\end{equation}
has been shown to define the field of a conformal string solution 
according to,
\begin{equation}
ds^2={\mathrm d}u{\mathrm d}v +  K(\chi) {\mathrm d}u^2 + {\mathrm d}\chi^2
+\sin^2\chi\left( {\mathrm d}\theta^2 +\sin^2\theta 
{\mathrm d}\varphi^2\right).
\label{Horus1}
\end{equation}
For all these reasons,  
the cotangent function presents itself as suited 
for playing the part of a conformal potential on 
the $S^3$ space of finite volume.
This curved space potential should not be confused with
the flat space Wilson loop potential generated by  $K(\chi)$  in the
surrounding infinite {\bf R}$^{1+3}$ space.\\ 

\noindent 
A further and independent motivation in favor of
employing the cotangent potential in quark models 
is provided by the observation \cite{CK_07} that
the lowest terms in  its Taylor expansion
coincide with a Coulombic+linear (Cornell) potential. This is easiest to
illustrate by the simplistic  $\chi=\frac{r}{R}\pi $ parametrization
(commonly used in super-symmetric quantum mechanics) for which 
\begin{equation}
-\cot \frac{r}{R}\pi =
-\frac{d}{r}+\frac{1}{3}\frac{r}{d} +
\frac{r^3}{45d^3}+
\frac{2r ^5}{945 d^3}+..., \quad 
\mbox{with}\quad  d=\frac{R}{\pi },
\label{crnl}
\end{equation}
holds valid.
The Cornell potential \cite{Cornell} has been
predicted by lattice QCD simulations \cite{Lattice}, on the one side,
and  has been also independently confirmed within the AdS/CFT context where
it emerges as a soft-wall Wilson loop potential
\cite{Wilson},\cite{Wen}. Its inverse distance part is associated with the
one gluon exchange of the perturbative regime, while the linear term 
is viewed to describe the flux-tube interactions of 
the non-perturbative regime. In order to approach the regime
of the asymptotic freedom, one needs to extend the Cornell potential
by corrections that account for more complicated non-perturbative processes. 
Such have been systematically explored within the topological approach in 
refs.~\cite{Brambilla}.
Within this context, the terms in eq.~(\ref{crnl})  beyond the first two
could  be viewed as a particular phenomenological 
parametrization of  non-perturbative corrections beyond the flux-tube
mechanism. As an advantage of such a parametrization we wish to mention
the exact solubility of the resulting potential.\\

\begin{quote}
Therefore, the cotangent potential on $S^3$, 
besides being congruent with conformal symmetry in
AdS$_5$/CFT$_4$,  also adequately  captures the dynamical aspects of 
QCD, a twofold advantage that  makes it attractive to 
applications in  hadron spectroscopy.
\end{quote}

\noindent
We shall introduce this very potential  as a gauge
interaction in eq.~(\ref{lpls_5})
by means of the replacement, 
\begin{eqnarray}
i\hbar \sqrt{\kappa} \frac{\partial }{\partial \tau} &\to &
i\hbar \sqrt{ \kappa} \frac{\partial }{\partial \tau} + 
2G\sqrt{\kappa }\cot \chi -{\bar a},\nonumber\\
\label{intr_intr}
\end{eqnarray}
where we parametrized $m$ in eq.~(\ref{Horus2}) as $m=-2G\sqrt{\kappa}$.
Upon factorizing the $\tau $ dependence of the total wave function
as $\exp \left(-i \frac{E}{\hbar \sqrt{\kappa}}\tau\right)  $, and after 
some algebraic manipulations, 
the interacting Klein-Gordon 
equation can be cast into the following form,
\begin{eqnarray}
{\Big(}-\kappa {\hbar^2}\frac{\mbox{d}^2 }{\mbox{d}\chi ^2} 
+ U_l(\kappa, \chi )
 & - &2G \sqrt{\kappa}(2E-2{\bar a})  \cot\chi
-(2G\sqrt{\kappa})^2 \csc^2\chi
 {\Big)}\Psi (\chi ) \nonumber\\
&=&
\lbrack (E-{\bar a})^2-\mu^2 
- \left( {\bar a}^2+ (2G\sqrt{\kappa})^2\right)\rbrack \Psi (\chi) . 
\label{chi_eq_intr}
\end{eqnarray}
The second  term on the l.h.s of this equation,
\begin{equation}
U_l(\chi, \kappa)= h^2\kappa l(l+1)\csc^2\chi,
\label{s3_brr}
\end{equation}
is the centrifugal barrier of the free geodesic motion on $S^3$.
Equation~(\ref{chi_eq_intr})  has been obtained in
making use of the peculiarity  
of the cotangent function to reproduce itself, and
the $S^3$ centrifugal barrier, upon squaring,
\begin{equation}
({\bar a}+m\cot \chi)^2 ={\bar a}^2 -m^2 +2{\bar a}m\cot \chi +m^2\csc^2\chi.
\end{equation}
Therefore, the  $\csc^2 $ terms arising 
upon the substitution of eqs.~(\ref{intr_intr}) in
eq.~(\ref{lpls_5}) can be absorbed by the $S^3$ centrifugal barrier.
Indeed, introducing the new constant,
\begin{eqnarray}
\alpha (l) &=&
-\frac{1}{2} +\sqrt{
\left( l+\frac{1}{2}\right)^2
-\frac{(2G)^2}{\hbar^2}}
= l+\Delta \, l,\nonumber\\
\Delta\, l&\approx & -\frac{1}{2}\frac{(2G)^2}{\hbar^2}\frac{1}
{l+\frac{1}{2}},
\label{new_a}
\end{eqnarray}
allows one to rewrite eq.~(\ref{chi_eq_intr}) equivalently to,
\begin{eqnarray}
 \lbrack  -\kappa {\hbar^2}\frac{\mbox{d}^2 }{\mbox{d}\chi ^2} 
+{\mathcal V}(\chi )\rbrack \Psi (\chi )&=&
((E-{\bar a})^2- c_0)\Psi (\chi),  \nonumber\\
{\mathcal V}(\chi )=
-2\beta \cot\chi  + {\overline U}_l(\chi, \kappa ) , &\quad &
{\overline U}_l(\chi, \kappa )=\hbar^2 \kappa \alpha (l) 
(\alpha (l) +1) 
\csc^2\chi, \nonumber\\ 
\beta = 2G \sqrt{\kappa}(E-{\bar a}),&\quad&
c_0= \mu^2 -\hbar^2\kappa + {\bar a}^2 + (2G\sqrt{\kappa})^2.
\label{chi_ctgnt}
\end{eqnarray}
The positive sign in front of the square root in eq.~(\ref{new_a}) 
ensures that ${\overline U}_l(\chi,\kappa)$
approaches  the $S^3$ centrifugal barrier of the free geodesic motion
in the $G\to 0$ limit, 
\begin{equation}
{\overline U}_l(\chi,\kappa )
\stackrel{G\to 0 }{\longrightarrow}U_l(\chi, \kappa).
\label{intr_brr}
\end{equation}

\noindent
Upon a suitable variable change, differential equations of the type
in (\ref{chi_ctgnt}) have been shown in refs.~\cite{CK_06}, \cite{raposo}
to reduce to one of the
forms of the hyper-geometric
equation whose solutions can be  expressed in closed form 
in terms of the non-classical Romanovski polynomials
(here denoted by $R_n^{(a,b)}(\cot \chi )$) .

\subsection{Wave functions, spectrum  and kinetic fine 
structure splittings }
The resulting explicit formula
for the wave functions then emerges as,
\begin{eqnarray}
\Psi_{Kl} (\chi)&=&
N_{Kl } e^{-a {\chi}}(\sin \chi )^{K+1+ \Delta l}
R_n^{(a,b)}(\cot \chi),
\label{psil} \\
a= \frac{2G(E-{\bar a})}{\sqrt{\kappa}\hbar^2 (K+1 +\Delta l)}, &\quad&
b=-(K+1 +\Delta l ), \quad K=n+l,
\label{constants}
\end{eqnarray}
where $N_{Kl}$  are normalization constants.
Correspondingly, the algebraic equation for the
energy  takes the form,
\begin{eqnarray}
(E-{\bar a})^2 &=&\frac{c_0+ {\hbar^2}\kappa (K+1+\Delta l)^2
}
{1 +\frac{4G^2 }{\hbar^2(K+1+\Delta l)^2}}.
\label{mss_frml}
\end{eqnarray}
The expansion of the latter equation to leading order in $\Delta l$ 
reads,
\begin{eqnarray}
(E-{\bar a})^2&\approx&
\left( E_{(K+1)}-{\bar a}\right)^2 
+ \Delta E_{(K+1)}^{(1)}(\Delta l), 
\label{expansion}
\end{eqnarray}
where the $l$ independent piece,
\begin{eqnarray}
(E_{(K+1)}-{\bar a})^2
&=&
\frac{
c_0+ \hbar^2\kappa (K+1 )^2
}
{
1 +\frac{
4G^2
}
{\hbar^2 (K+1)^2}
},
\end{eqnarray}
can be viewed as an unperturbed  degeneracy energy.
The difference, $(E-{\bar a})^2-(E_{(K+1)}-{\bar a})^2=
\Delta E_{ (K+1)}^ {(1)}
(\Delta l)$, 
then calculates as
\begin{eqnarray}
\Delta E_{ (K+1)}^ {(1)}
(\Delta l)&=&2\Delta l
\left( \frac{\hbar^2\kappa (K+1)}{ 1+\frac{4G^2}{\hbar^2(K+1)^2}}
-\frac{4G^2\left( \frac{c_0}{(K+1)^3}+\frac{\hbar^2\kappa }
{K+1}\right)}{\hbar^2 
\left( 1+\frac{4G^2}{\hbar^2(K+1)^2}\right)^2}
\right).
\end{eqnarray}
As long as the ``interacting''  principal quantum number, $(K+1 +\Delta l)$,  
in eq.~(\ref{mss_frml}) can be at most close to integer,
the degeneracy of states within the $(K/2,K/2)$
multiplet in eq.~(\ref{Casimir_O4}) is relaxed.
Nonetheless, as visible from eq.~(\ref{expansion})
the spectrum is still patterned after the $SO(4)$ levels $|Klm>$,
and it still falls into an 
$\infty$d unitary representation of
the conformal group. As we shall see below,
the $\Delta l $ contributions will have a detectable effect only  
on the masses of states with $l=0$ and will be
helpful  in removing the degeneracy between
$P_{1/2}$--$S_{1/2}$ states,
while leaving all the other excitations  
practically degenerate. The latter property of the interacting conformal
rigid rotor parallels on $S^3$ the kinetic fine level splittings of a 
hydrogenic two-body system in a plane space and will 
occasionally be referred to as ``curved'' fine level splitting.

In effect, the relativistic  framework of 
the conformal scale equation~(\ref{conf_eq}) on
{\bf R}$^1\otimes S^3$, gauged (\ref{chi_eq_intr}), 
provides the 
intriguing possibility of having exclusively bound states
organized into  conformal bands whose states are 
not necessarily perfectly degenerate
though they still keep spreading around $SO(4)$ levels.
The case in which one of the particles is a spin-1/2 fermion is 
easily incorporated into the formalism by coupling
the Dirac spinor (1/2,0)$\oplus$(0,1/2) to $|Klm>$ in eq.~(\ref{Casimir_O4}).
The excitations of such a system populate an infinite staircase
(that is an $\infty$d unitary SO(2,4) representation) whose ladders
are given by reducible SO(4) representations as,
\begin{equation} 
\left( \frac{K}{2}, \frac{K}{2}\right)\otimes \left[ 
\left( \frac{1}{2},0\right) \oplus \left(0,\frac{1}{2} \right)\right],\quad
K=0,1,2,..\infty.
\label{sdof}
\end{equation}
Such multiplets consist of  $K$ parity dyads of rising spins, 
$j^P =\frac{1}{2}^\pm ,...,\left(K-\frac{1}{2}\right)^\pm $, 
and a single-parity 
state of maximal spin, $j^P=\left(K+\frac{1}{2}\right)^P$. 
The absolute value of the parity of the two-body system 
depends on the parity of the scalar body,
$\pi$, and is given by $P=\pi (-1)^l$.  
Alternatively, one also could have started from the very beginning with   
the non-relativistic  stationary Schr\"odinger equation on $S^3$ with
the $\cot\chi$ potential as                      
\begin{eqnarray}
{\Big[} \frac{\hbar^2}{2 m }\kappa {\mathcal K}^2 -2G\kappa \cot\chi 
&-& {\mathcal E}{\Big]} 
\psi  (\chi )=0,
\label{chi_eq_lin}
\end{eqnarray}
an option  first considered 
by Schr\"odinger \cite{Schr40}.        

This option has been investigated by us in our prior works 
\cite{CK_06}, \cite{CK_07}. 
In contrast to eq.~(\ref{chi_eq_intr}), the linear in
mass Schr\"odinger equation (\ref{chi_eq_lin}) keeps
respecting in the interacting case the degeneracies of the free
geodesic motion \cite{CK_09}. 
In the present work we systematically departure from eq.~(\ref{chi_eq_intr}).
We shall compare outcomes of these two 
schemes in due places.\\

\section{Quark-diquark model on $S^3$. }
From now onward  we assume dominance of quark-diquark configurations
in the internal nucleon and  $\Delta$ structures and apply 
eq.~(\ref{chi_eq_intr}) to  the description of the relative motion
of these two bodies. We moreover shall consider the diquark as spin-less, 
a restriction that enables one to hit a specifically simple  
unitary $SO(2,4)$ representation. As we shall see below, 
this configuration turns out to be the  
quite adequate for data description.
In considering the diquark as spinless,
the total spin, $J$,  of the resonance is then obtained
through coupling the spin of the quark to the $q-(qq)$ 
relative angular momentum, $l$.
For the time being, and because of the absence of spin-flavor 
interactions in the
wave equation under consideration,
we shall factorize the light flavor quantum number
(actually isospin).
In due course we shall see that the spectra reported so far
do not contradict  the above assumptions. 
The application of the conformal 
equation~(\ref{conf_eq}) to such two-body systems 
is then  straightforward and can be featured as follows: 

\begin{itemize}
\item The relative angular momentum between the
quark and the spin-less diquark takes the values $ l\in[0,K]$,
in accord  with  the conformal branching rule connecting the
${\mathcal K}^2$, and  {\bf L}$^2$ eigenvalues 
in eq.~(\ref{Casimir_O4}).

\item The total spin, $J$,  of the $q$--$(qq)$ system
is obtained from coupling the quark spin to that very $l$ as
\begin{equation}
\lbrack {\vec l}
\otimes{ \vec {\frac{1}{2}}}
\rbrack^{JM},
\quad J=l\pm \frac{1}{2}.
\label{coupling_scheme}
\end{equation}

\item  In effect, one finds the
following fermionic $SO(4)$ multiplets,
\begin{eqnarray}
J &\in& 
\left( \frac{K}{2}, \frac{K}{2}\right)\otimes \left[
\left(\frac{1}{2},0\right)\oplus \left(0, \frac{1}{2} \right)\right],
\quad l\in [0,K], 
\label{degenrcy}
\end{eqnarray}
emerging as levels of a conformal band corresponding to,
\begin{equation}
\quad K\in [0,\infty ).
\end{equation}
Accordingly, the total spin $J$ in eq.~(\ref{coupling_scheme}) 
takes the following values,
\begin{eqnarray}
J&=&\frac{1}{2}, \left( 1\pm \frac{1}{2}\right), 
\left( 2\pm \frac{1}{2}\right), ...,
\left( K\pm \frac{1}{2}\right),
\label{degenrcy_1}
\end{eqnarray}
to be identified with the spins of the light flavor baryon resonances. 

\item The parity, $P$, of the resonances, $P=\pi(-1)^l$,
is the product of the parity, $\pi=\pm $, of the spinless diquark, 
and $(-1)^l$, the parity of the relative angular motion.
The parity of the diquark, scalar or pseudoscalar, is fixed by 
matching parity, $P_{\mbox{\footnotesize max}}$, 
of the highest spin, $J=K+\frac{1}{2}$ ,
i.e. 
\begin{equation}
\mbox{If}\quad J^{P}=
\left( 
K+\frac{1}{2}
\right)^
{
P_{\mbox{\footnotesize max}}
}, \quad \mbox{then} \quad  \pi =P_{\mbox{\footnotesize max}} (-1)^K. 
\label{party}
\end{equation}
\item  As a working hypothesis, to be tested by comparison with data,
the nucleon diquark has been set as a scalar  
for the ground state and the low lying excitations with masses below 
~1600 MeV, and as a pseudo-scalar  above.

\item The $\Delta$ diquark has been set as an axial vector in the 
ground state and as a pseudo-scalar above.

\item The pseudo-scalar  diquark 
is a $P$ wave axial-vector  whose total angular momentum equals zero,
and emerges from coupling the axial spin vector, ${\vec S}={\vec 1}^+$, 
to an internal vectorial, $1^-$, excitation according to
$\lbrack {\vec 1}^+\otimes {\vec  1}^-\rbrack ^{0^-}$.

\item The quark-diquark system is described by means of 
eq.~(\ref{chi_ctgnt}) with the wave functions in eq.~(\ref{constants}).
Insertion of the $\chi$ parametrization from eq.~(\ref{RW_prmtrz}) into
eq.~(\ref{chi_ctgnt}) amounts to the following spatial 
$q$--$(qq)$ wave function,
\begin{eqnarray}
\Psi_{Kl} \left(\chi (r)\right)&=&
N_{Kl } e^{-a {\sin^{-1}r\sqrt{\kappa}}}(r\sqrt{\kappa})^{K+1+ \Delta l}
R_{K-l}^{(a,b)}(\cot \sin^{-1}r\sqrt{\kappa}).
\label{psi_rad} 
\end{eqnarray} 
It represents the radial dependence of the total wave function
within the finite volume position space,
\begin{equation}
r\sqrt{\kappa}\in [0,1].
\label{fnt_cnft}
\end{equation}

\item The energies $E$ in eq.~(\ref{mss_frml}),   
will be subsequently re-denoted by $M$ and
given interpretation of resonance masses read off
from the respective ground state $N$ and $\Delta$ masses.

\end{itemize}

In this fashion, the spatial part of the
total baryon wave function has been designed to account for
conformal symmetry in accord with AdS$_5$/CFT$_4$ on the one side, and
with the (approximate) conformal symmetry of the QCD Lagrangian in the
light flavor sector, on the other. In addition, 
$\Psi_{Kl}\left( \chi (r)\right) $ describes a confinement of finite range.  
In the next section we shall compare the outcome of such a model
with data on $N$ and $\Delta$ resonances.

\subsection{The N and $\Delta$  spectra}

The spectrum of the nucleon continues being under debate despite 
the long history of the respective studies ~\cite{Lee}, 
\cite{Afonin}.
Yet, unprejudiced inspection of the data reported by the Particle
Data Group \cite{PART} reveals systematic and hardly to overlook
grouping of the
excited states of the baryons of the best coverage, the
nucleon, and the $\Delta (1232)$. 
Take as a prominent example the seven $\Delta $ resonances 
$S_{31}(1900)$, $P_{31}(1910)$, $P_{33}(1920)$,$D_{33}(1940)$,
$F_{35}(1905)$,   $D_{35}(1930)$, and $F_{37}(1950)$, which are
squeezed in the narrow mass band between 1900 MeV to 1950 MeV
and which, given the limitations  of data accuracies,
are practically mass degenerate.
This sequence of resonances consists of 3 parity dyads with spins ranging
from $\frac{1}{2}^\pm$ to $\frac{5}{2}^\pm$ and of a single parity state
of maximal spin, $J^\pi =\frac{7}{2}^+$ and its quantum numbers fit
into the $K=3$ multiplet in eqs.~(\ref{degenrcy}),(\ref{degenrcy_1}) .
In the nucleon spectrum, one finds the 
$N(1440)$--$N(1535)$--$N(1520)$ triplet, which 
would match quantum numbers of $K=1$ in eq.~(\ref{degenrcy}).
Patterns of similar type do not restrict to these two examples 
alone but extend to the spectrum  of any of the lightest 
flavor baryons, no matter $N$, or $\Delta$.
Notice the following observations.
\begin{itemize}
\item The $\Delta (1232)$ excitations around 1700 MeV  
form a group of three that contains  one parity dyad with lowest 
spin $\frac{1}{2}^\pm$ and one single parity state of maximal spin, 
$J^\pi =\frac{3}{2}^-$ and would match  $K=1$ in eq.~(\ref{degenrcy}),
\item The $\Delta$ resonances around 1900 MeV are just
the $K=3$ example from above,
which consisted of the three mass-degenerate parity dyads with 
spins ranging form
$\frac{1}{2}^\pm$ to $\frac{5}{2}^\pm$, and the one 
unpaired state of maximal spin, 
$\frac{7}{2}^+$, of practically  same mass.
\item Another spin-sequence could match $K=5$ in eq.~(\ref{degenrcy})
and  is well marked
by the parity simplex $J^\pi =\frac{11}{2}^+$ of maximal spin in this region, 
the four star resonance $H_{3,11}(2420)$.
The pairs $H_{39}(2300)$--$G_{39}(2400)$, 
and $G_{37}(2200)$--$F_{37}(2390)$ would
provide two of the
five parity dyads necessary to complete $K=5$ in eq.~(\ref{degenrcy}).
The $D_{35}(2350)$ would be suited as a part of a $\frac{5}{2}^\pm$ doublet,
while spin $\frac{1}{2}^\pm$ and $\frac{3}{2}^\pm$ pairs are ``missing''. 
\end{itemize}
It verifies directly by inspection  that the mass-splittings 
inside the above spin-cascades are notably smaller than the splittings 
between the averaged  cascade masses. In  this fashion  the $\Delta $ spectrum
reveals a level structure.
The $N$ excitations follow quite same patterns though  appear shifted 
downward in mass by about 200 MeV relative to the $\Delta$  excitations.
The attention to this clustering phenomenon in light baryon spectra
has been drawn first in ref.~\cite{MK_97}
and was further elaborated in refs.~\cite{MK_2000}-\cite{MK_Evora}.
In assuming equality of the quantum numbers of the $N$ and $\Delta$ 
excitations and comparing both spectra, allows to pin down the states
missing for the completeness of the scheme.

\noindent
Under the assumption of  $q-(qq)^{0^\pm }$
as a dominant  configuration of internal baryon structure,
the above clustering phenomenon is shown below to
find a quantitative explanation in terms of the levels of 
the spectrum of 
eqs.~(\ref{chi_ctgnt}),~(\ref{mss_frml}),~(\ref{degenrcy}), and 
(\ref{degenrcy_1}), and is
interpreted here as a hint on possible realization
of conformal symmetry 
in the light flavor baryon spectra.\\ 

\noindent
\underline{Missing resonance predictions.}\\

\noindent
The  comparison of the spectrum in eq.~(\ref{mss_frml})
to data \cite{PART}
is presented in Figs.~\ref{levels-Delta}, 
and~\ref{levels-Nucleon}. 
\begin{figure}
\resizebox{0.80\textwidth}{7.5cm}
{\includegraphics{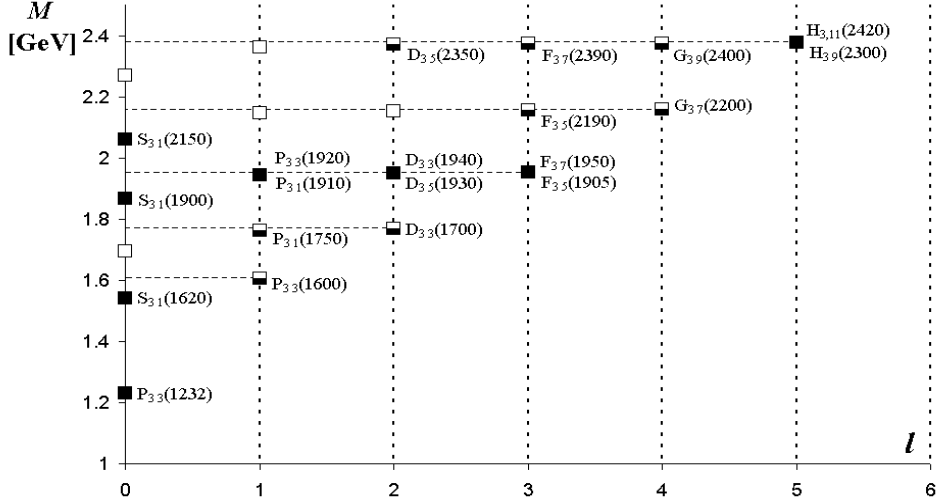}}
\caption{
Assignments on the $l/M$ grid 
of the reported \cite{PART} $\Delta $ excitations to
the {\bf R$^1\otimes S^3$} spectrum in eq.~(\ref{mss_frml}). 
The distribution of the reported (exp) resonances over the
predicted (th) states  has been obtained from a
least mean square four parameter data fit, i.e. by minimizing
$\sigma=
\sqrt{
\frac{1}{N}
\Sigma_{i=1}^{i=N} 
\left(
M^{(i)}_{
\mbox{\footnotesize th}}
-M^{(i)}_{\mbox{\footnotesize exp }} 
\right)^2 }$. 
The sum includes all the reported resonances.
The minimal value,
$\sigma_{\mbox{\footnotesize min}}= 0.043 \, 
\mbox{GeV}$ 
has been obtained for the following
potential parameters: $G=0.04933$ GeV$\cdot$fm, $R=0.747$ fm,
${\bar a}=0.5037$ GeV, and 
$\mu =1.044$ GeV.
The excitations, $E$, set equal to masses,
 have been read off from the $\Delta (1232)$ mass.
Full and empty squares denote reported and predicted states, respectively.
\label{levels-Delta}}
\end{figure}
\begin{figure}
\resizebox{0.80\textwidth}{7.5cm}
{\includegraphics{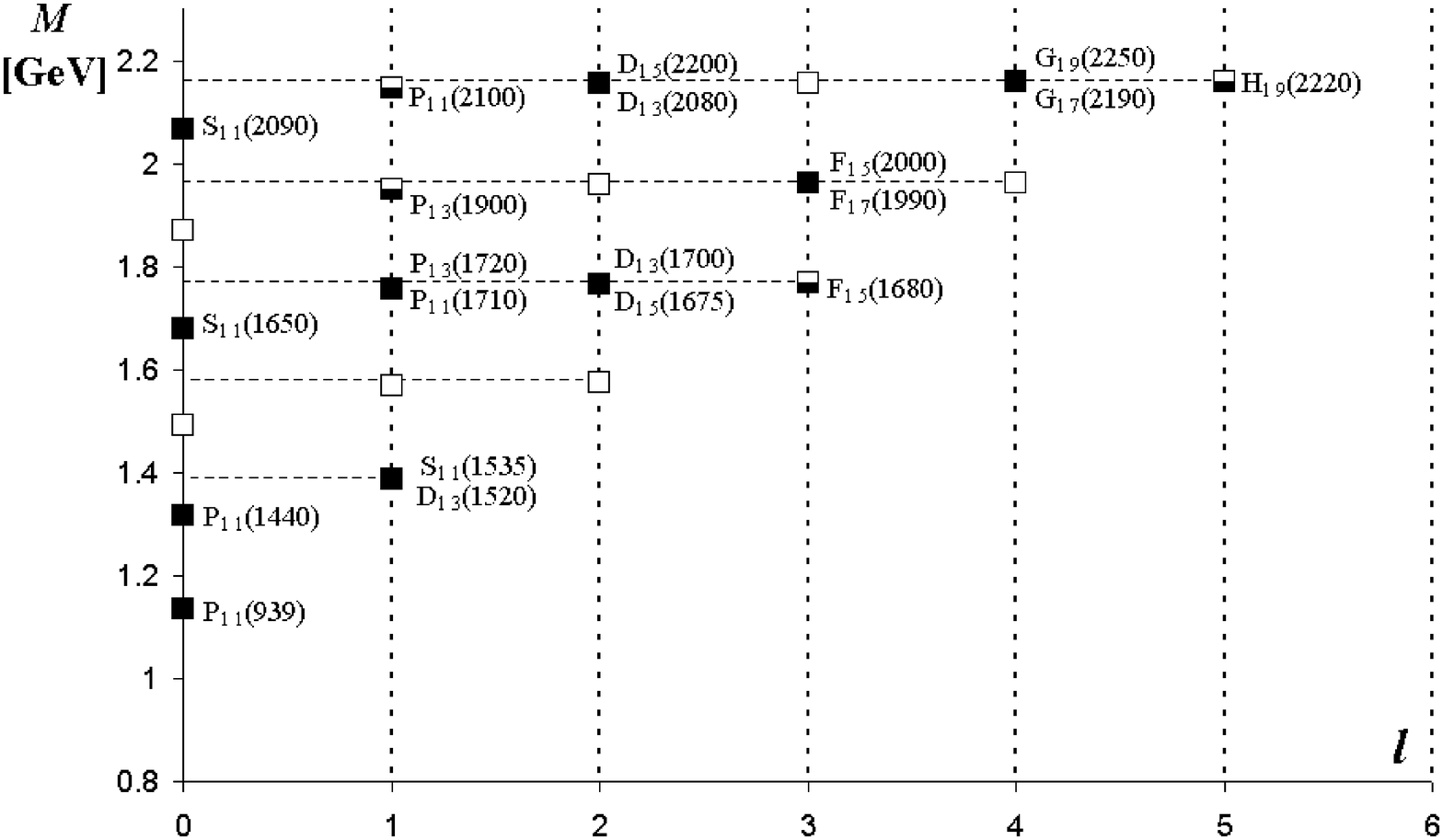}}
\caption{
Assignments on the $l/M$ grid 
of the reported \cite{PART} $N$ excitations to
the {\bf R$^1\otimes S^3$} spectrum in eq.~(\ref{mss_frml}).
The distribution of the reported (exp) resonances over the
predicted (th) states has been obtained from a
four parameter least mean square data fit, i.e. by minimizing
$\sigma=\sqrt{\frac{1}{N} \Sigma_{i=1}^{i=N} 
\left( M^{(i)}_{\mbox{\footnotesize th}}\, 
-M^{(i)}_{\mbox{\footnotesize exp }}\, \right)^2 }$.
The sum includes all the reported resonances.
The value of the proton mean square charge radius has been also
taken into consideration by the fit.
The minimal value of
$\sigma_{
\mbox{\footnotesize min}}=0.0855$ GeV 
has been obtained for
the following  potential parameters: 
$G= 0.0493$ GeV$\cdot $fm, $R=0.9814$ fm, and
$\mu =0.3213$ GeV, and ${\bar a}=0.932$ GeV.
The excitations, $E$, set equal to masses, 
have been read off from the  nucleon  
mass.
Other notations as in Fig.~\ref{levels-Delta}
\label{levels-Nucleon}}
\end{figure}

\noindent
We predict one $P_{31}$ ``missing'' state from
the $K=1$ level. The three more  states, $S_{31}$, $P_{33}$, and $D_{35}$
are  needed for the completeness of the $K=2$ level.
The $K=3$ level is complete. From the required nine states in the
next $K=4$ level, $S_{31}$, and $G_{37}$, have been
in turn identified with the observed one star resonances
$\Delta (2150)$, and $\Delta (2200)$.
Our $F_{35}$ has been
associated with the two-star resonance
 $\Delta (2000)$, for which the two quite different mass values of
$(1752\pm 32)$, and $(2200\pm 125)$ have been
listed in \cite{PART}. We here adopt the position of ref.~\cite{dong} and
consider this resonance as $\Delta (2190)$, which allows to place it
well  in the $K=4$ multiplet.
 The remaining six states from that very level are missing (see Table 
\ref{best_fit_Dlt}.)
Finally, the  $K=5$ members  $H_{39}$, and $G_{39}$ can be assigned to
the respective two-star resonances $\Delta (2300)$, and $\Delta (2400)$, while
$\Delta (2350)$ and $\Delta (2390)$ are good candidates for
$D_{35}$,  and $F_{37}$. This level is well marked by its highest spin
$H_{3,11}$, the four-star resonance $\Delta (2420)$.
We find 16 resonances missing from the first five levels of the
the conformal band covering the $\Delta$ spectrum.
On the nucleon side the highest spins , 
$F_{17}$ from the $K=3$,  and $H_{31, 11}$ from 
$K=5$  are ``missing'' . The K=2 level is  completely 
``missing''. 
Compared to our previous work \cite{CK_06},\cite{CK_07}, the
fit places the N(1900), N(1990), and N(2000) resonances at the
lower $K=4$ level and leaves instead $ P_{11},
F_{15},$ and $ F_{17}$ in K=5 unoccupied.
The number of ``missing'' nucleonic states is sixteen
(see Table \ref{best_fit_Dlt}). 
Therefore, for baryons whose internal structure is dominated by the
$q$--$(qq)^{0^\pm}$ configuration, 
we predict a total of 32
nucleon and $\Delta$  resonances ``missing'' from
the first five levels of the
respective nucleon, and $\Delta$ conformal bands.
Figures and tables show that the $N$ and $\Delta$ resonances reported so far
are pretty well matched by the excitations of this simplest configuration  
and are illustrative of a well
pronounced  footprint of conformal symmetry in the spectra of the
lightest flavor baryons.
In case the diquarks were to carry higher angular
momenta, the  excitations  of $q$--$(qq)^{l}$ with $l>0$ 
would populate higher $SO(2,4)$ 
representations, 
which one can expect to appear much heavier in mass.   
One of the 
conclusions following from our findings is that at most 
 ``missing'' states belonging to different levels of the 
conformal band  might still have chances
to be pinned down by an $ O(3)$ partial wave analysis.
In contrast, the states belonging to same level, in being strongly overlapping,
are better looked up in terms of $O(4)$ partial wave analysis and 
identified as a  whole.

In conclusion, the confirmation by data of the
predicted degeneracy among the parity pairs belonging to same
$K$ and displayed in Figs.~ 1 and 2,
signals relevance of conformal symmetry for the
spectra of the light flavor baryons.\\

\noindent
Before proceeding further, a  comment is in order  on
the degeneracies predicted by the Light-Front QCD framework 
\cite{Br1}, \cite{Br2}.
There, one finds the mass formula as
\begin{equation}
M^2=N, \quad N=1,2,3,..., \quad N= n+\nu +1,
\label{LC1}
\end{equation}
and observes  again a conformal band (as it should be)
with respect to $\nu$. The relation of the light-cone variable
$\nu$  to ordinary angular momentum is more involved. It equals
 $\nu=L$ for mesons, and $\nu=L+1$ for baryons.
Also the Light-Front QCD formalism reports degeneracies of excited
$N$ and $\Delta $ states, that partly overlap with those
considered here. Examples are the positive
parity spin-sequences F$_{37}$(1950)--
F$_{35}$(1905)--P$_{33}$(1920)--P$_{31}$(1910) from the
$K=3$ level in Fig.~\ref{levels-Delta}, and
the F$_{15}$(1680)--P$_{13}$(1720) states from 
same level in the nucleon spectrum in
Fig.~\ref{levels-Nucleon}. So far, the question on the
degeneracies among opposite parities has not been addressed 
in refs.~\cite{Br1}, \cite{Br2}.\\

\noindent
\underline{Mass ordering in $P_{2I,1}$--$S_{2I,1}$ pairs.}\\

\noindent
The model predicts the correct mass ordering of the 
$P_{11}-S_{11}$ states through
the spectrum. 
Within the framework of the present study the numerical value of
the splittings between such states is entirely determined by the
gauged  centrifugal barrier, ${\overline U}_l(\chi, \kappa )$,
defined in eqs.~(\ref{chi_ctgnt}),
which prescribes that $l=1$ states will  appear higher in mass than 
those with $l=0$. The ordering, $P_{2I,1}$--$S_{2I,1}$ versus
$S_{2I,1}$--$P_{2I,1}$ depends on the parity of the diquark.
When the diquark is a scalar, spin-$\frac{1}{2}^+$ and 
spin-$\frac{1}{2}^-$ in turn refer to 
zero and unit underlying angular momenta and 
are associated with $P_{2I,1}$, and  $S_{2I,1}$
states. This is the reason for which at the scale of  1500 MeV,  
where the diquark is a scalar,
the measured  $P_{11}(1440)$ state appears  lower in mass
than its $S_{11}(1535)$ neighbor.  From
the 1700 MeV level onward, the parity of the diquark changes
to pseudoscalar, and it is 
$S_{11}$ that has zero angular momentum, 
in accord to eq.~(\ref{party}).
Consequently,  $S_{11}$ states with masses above 1600 MeV
appear systematically at lower masses
than their nearest  $P_{11}$ neighbors. Examples are the
$S_{11}(1650)-P_{11}(1710)$, and $S_{11}(2090)-P_{11}(2100)$ pairs.
Recall that originally the suggestion on the parity change of the
diquark  was made with the purpose
of  matching parity of the highest spins in the fermionic multiplets
in eqs.~(\ref{degenrcy_1}), (\ref{party}).
 Therefore,
\begin{quote}
the reverse mass ordering in the $S_{11}-P_{11}$ pairs 
above 1600 MeV relative to the $P_{11}(1440)-S_{11}(1535)$ 
splitting, provides an independent argument
in favor of the change of parity of the diquark from 
scalar to pseudoscalar at that scale.
\end{quote}

\noindent
In the $\Delta$ spectrum,
where the diquark was fixed to a pseudoscalar for
all the excited states, we find  similar 
$S_{31}$--$P_{31}$ splittings.  The mass ordering in the
$S_{31}(1900)$-$P_{31}(1910)$ pair, also correctly described within 
the model presented, serves as an example.
As to the resonances with $l>1$, the $\Delta l$ corrections to $l$
in eq.~(\ref{new_a}) become insignificant and the splittings practically
disappear (see Tables \ref{best_fit_Dlt} and \ref{best_fit_Ncl}).\\

\noindent
In conclusion, we observed that the
quantum numbers of the reported nucleon resonances are
close to equal to those of the
reported $\Delta $ resonances (compare Tables \ref{best_fit_Dlt}, and 
\ref{best_fit_Ncl}).
We took advantage of this circumstance to
embed the reported states isospin by isospin
into conformal bands
of the type in eq.~(\ref{degenrcy}).
In so doing we patterned the $N$, and the $\Delta$ spectrum each
after an unitary representation of the conformal group.  
So far, no reported state drops out of the 
suggested systematics. We predicted a  total of
32 ``missing'' resonances 
needed for the completeness of the conformal nucleon and 
$\Delta$ bands. 
\begin{table}
\vspace*{0.21truein}
\caption{ Predicted excitation values  of the $\Delta$  states
in the $K=0\div 5$ levels of the conformal band in eq.~(\ref{Casimir_O4})
calculated according to eq.~(\ref{mss_frml}). The 
internal angular momenta take the values, $l=0,1,2,..K$.  ``Missing'' 
states are marked by boldface. 
The number in the parenthesis gives the predicted mass of excitations [in MeV]
carrying same internal angular momentum. 
States with $J=l\pm \frac{1}{2}$ appear degenerate
within the present scheme 
because of the absence of spin-orbit interactions in the present stage model.
Notice that differently from its excitations,
$\Delta (1232)$  belongs to
$\left(\frac{3}{2},0 \right)\oplus \left(0, \frac{3}{2} \right)$. }
\vspace{0.21truein}
\begin{tabular}{llccccc}
\hline
~\\
K&l=0  & l=1& l=2
& l=3 &l=4&l=5 \\  
~\\  
\hline
\hline
~\\
K=0& P $_{31}$(1230) & -& -& -& -&-\\
\hline
~\\
K=1 &S$_{31}$(1542)&{\bf P}$_{\mathbf 31}$/P$_{33}$(1607)&-&-&-&\\
\hline
~\\
K=2&{\bf S}$_{\mathbf 31}$(1699) &P$_{31}$/{\bf P}$_{33}$( 1768) &
D$_{33}$/{\bf D}$_{\mathbf 35}$(1774)  
&- & -&-\\
\hline
~\\
K=3&S$_{31}$(1874)&P$_{31}$/P$_{33}$(1953) &D$_{33}$/D$_{35}$(1960)  &
F$_{35}$/F$_{37}$(1963) &- &- \\
\hline
~\\
K=4&S$_{31}$(2072)&{\bf P}$_{\mathbf 3 1}$/{\bf P}$_{\mathbf 33}$(2159) &
{\bf D}$_{\mathbf 3 3}$/{\bf D}$_{\mathbf 3 5}$(2167 ) &
F$_{35}$/{\bf F}$_{\mathbf 37}$(2170) &G$_{37}$/{\bf G}$_{\mathbf 39}$(2171)&-  \\
\hline
~\\
K=5&{\bf S}$_{\mathbf 31}$(2287)&
{\bf P}$_{\mathbf 3 1}$/{\bf P}$_{\mathbf 33}$(2380) &{\bf D}$_{\mathbf 33}$/
D$_{35}$(2388)  &
{\bf F}$_{\mathbf 3 5}$/F$_{37}$(2391) &{\bf G}$_{\mathbf 37}$/G$_{39}$(2393)&
H$_{39}$/H$_{3,11}$(2394) \\
~\\
\hline
\end{tabular}
\label{best_fit_Dlt}
\end{table}

\begin{table}
\vspace*{0.21truein}
\caption{ Predicted excitation values of the nucleon states
in the $K$=$0\div 5$ levels of the conformal band in eq.~(\ref{Casimir_O4})
calculated in accord with eq.~(\ref{mss_frml}). Other notations as in 
Table 1. }
\vspace{0.21truein}
\begin{tabular}{llccccc}
\hline
~\\
K&l=0  & l=1& l=2
& l=3 &l=4&l=5 \\  
~\\  
\hline
\hline
~\\
K=0&P$_{11}$(1136)& -& -& -& -&-\\
\hline
~\\
K=1&P$_{11}$(1316)&S$_{11}$/D$_{13}$(1387)&-&-&-&\\
\hline
~\\
K=2&{\bf P}$_{{\mathbf 11}}$(1492) &
{\bf S}$_{\mathbf 11}$/{\bf D}$_{\mathbf 13}$( 1568) &
{\bf P}$_{\mathbf 13}$/{\bf F}$_{\mathbf 15}$(1574)  
&- & -&-\\
\hline
~\\
K=3&S$_{11}$(1678)&{ P}$_{ 11}$/P$_{13}$(1757) &D$_{13}$/{ D}$_{ 15}$(1764)  &
F$_{15}$/{\bf F}$_{{\mathbf 17}}$(1767) &- &- \\
\hline
~\\
K=4&{\bf S}$_{\mathbf 11}$(1870)&
{\bf P}$_{\mathbf 11}$/ P$_{ 13}$(1951) &{\bf D}$_{\mathbf 13}
$/{\bf D}$_{\mathbf 15}$(1958 ) &
{ F}$_{15}$/F$_{17}$(1961) &{\bf G}$_{\mathbf 17}$/{\bf G}$_{\mathbf 19}$(1962)
&-  \\
\hline
~\\
K=5&S$_{11}$(2066)&
P$_{11}$/{\bf P}$_{\mathbf 13}$(2147) &D$_{13}$/D$_{15}$(2154)  &
{\bf F}$_{\mathbf 15}$/{\bf F}$_{\mathbf 17}$(2157) &G$_{17}$/G$_{1,9}$(2159)&
H$_{1,9}$/{\bf H}$_{\mathbf 1,11}$(2160) \\
~\\
\hline
\end{tabular}
\label{best_fit_Ncl}
\end{table}

\subsection{Charge radii and form factors}

The effect of curvature on the physical observables is two-fold.
On the one side, it encodes the topology of the position space and
will influence Fourier transforms
through the $S^3$ integration volume, 
 \begin{equation}
{\mathrm d}\, \Omega_3= \sin^2\chi \sin\theta \mbox{d}\chi \mbox{d}
\theta \mbox{d}\varphi=
(r\sqrt{\kappa})^2\sin \theta {\mathrm d}\left( \sin^{-1}r\sqrt{\kappa}\right)
\,{\mathrm d}\theta {\mathrm d}\varphi ,
\label{4d_vlm}
\end{equation}
where use has been made from  the parametrization in eq.~(\ref{RW_prmtrz}).
The curved integration volume, 
$(r\sqrt{ \kappa } )^2 d\sin^{-1}r\sqrt{\kappa}$,
approaches the flat one, $\kappa^{\frac{3}{2}}r^2dr$, 
only in the small $\chi$ angle 
approximation. On the other side,  curvature  can be viewed as a
new phenomenological potential parameter.
The first aspect is of crucial importance in performing the
Fourier transform of the cotangent potential and constructing
an effective instantaneous gluon propagator, in parallel to the
instantaneous photon propagator obtained from Fourier transforming
the Coulomb potential.
Exploring this aspect is the subject of section 4.
Compared to this,  calculations of form-factors of states of low internal 
angular momentum, such as the
ground state, and the $P_{11}(1440)$ resonance, are much less affected by the
integration volume, because 
the wave functions practically correspond to  
the small $\chi$ angle approximation, and 
thereby approach flat-space wave functions. In these calculations the
importance of curvature is more
to provide a phenomenological parameter in addition to the potential
strength and facilitate data fits. \\

\noindent
The standard scheme for calculating  form-factors relies upon
Fourier transforms from position to the conjugate
momentum space by means of the 3d plane wave, 
$\exp\, i {\mathbf q}\cdot {\mathbf r}$. This plane wave 
can be regarded  as the special case of the {\bf R}$^4$  plane wave,
$\exp \, (i q_4x_4 + {\mathbf q}\cdot {\mathbf r})$, in which
$q_4=0$. From this perspective, the 3d plane wave that accounts for a 
position vector ${\mathbf r}$ restricted to the equatorial plane 
of $S^3$ reads,
\begin{equation}
e^{iq_4+i|\mathbf{q}||\mathbf{ r}|\cos \theta }|_{q_4=0}= 
e^{i|\mathbf{q}||\mathbf{ r}|\cos \theta }=
e^{i{|\mathbf{q}|}\frac{\sin \chi}{\sqrt{{\kappa}}} 
\cos \theta },
\quad |\mathbf{ r}|=R\sin\chi=\frac{\sin \chi}{\sqrt{{\kappa}}},
\quad r\sqrt{\kappa}\in [0,1],
\label{4_PW_FF}
\end{equation}
and refers to a $z$ axis chosen along the
momentum vector (a choice justified in elastic scattering). 

Electric charge form-factors are the simplest physical observables to 
calculate, and the corner stone of any spectroscopic model.
They reduce to the Fourier transform of the charge density, proportional
to $|\Psi_{Kl}(\chi )|^2$ in our case.  
The extraction of the mean square charge radius,
$<{\mathbf { r}}^2>$, from the form-factor
is then standard and calculated as the slope at origin.
We here choose as illustrative examples the mean square charge radii of the
proton, the $P_{11}(1440)$ and $S_{11}(1535)$ resonances.
The proton ground state wave function 
entering the calculation is obtained from eq.~(\ref{constants}), the
explicit expression for the normalization factor being
\begin{eqnarray}
N_{(00)}&=&\frac{4b(a ^2+1)}{1-e^{-2\pi a }},
\quad a=
\frac{2GE_0}{\hbar^2 \sqrt{\kappa }
\left(1 -\frac{1}{2}+
\sqrt{\frac{1}{4}-\frac{(2G)^2}{\hbar^2}}\right)
}, \quad E_0={\bar a} +\sqrt{\frac{1}{2}\mu^2 -\frac{\hbar }{4}\kappa }.
\label{wafu_gst}
\end{eqnarray}
The mean square charge radius for this state expresses
in closed  form as,
\begin{equation}
<{\mathbf { r}}^2>_p=
\frac{6}{ (4a^2+9) \sqrt{\kappa} }.
\label{r2_closed}
\end{equation}
With the potential parameters fitted to the spectra, we find,
\begin{equation}
\sqrt{<{\mathbf r^2}>_p}=0.664 \,\, \mbox{ fm},
\end{equation}
to be compared to the experimental value of
$\sqrt{<{\mathbf { r}}^2>_p}=0.8750 \pm 0.008$ fm reported by
\cite{PART}.
Comparison of the predicted to the measured proton
electric charge form factor is presented in Fig.~3.
For the Roper resonance, same observable is calculated 
as
\begin{equation}
<{\mathbf { r}}^2>_{\mbox{\footnotesize Roper }}=
\frac{2(5+52 a^2)}{(10 a^4 +104 a^2 +25)\kappa },
\quad a^2 =\frac{4G^2(E-{\bar a})^2}{\hbar^4\kappa (\alpha (l)+1)^2}.
\end{equation}
The resulting value is
\begin{equation}
\sqrt{<{\mathbf { r}}^2>_{\mbox{\footnotesize Roper}}}=
0.8484 \,\, \mbox{ fm}. 
\end{equation}
As to the $S_{11}(1535)$ resonances, the prediction is numerical
and obtained as  
\begin{equation}
\sqrt{<{\mathbf { r}}^2>_{
\mbox{
\footnotesize S}_{11}
}
}= 0.8754
 \,\, \mbox{ fm}. 
\end{equation}
The proton form factor is compared to data in Fig.~\ref{Form_F}.
The comparison of the other two form-factors to that of the proton 
is presented in Fig.~\ref{Formfs}.
\begin{figure}
\resizebox{0.80\textwidth}{7.5cm}
{\includegraphics{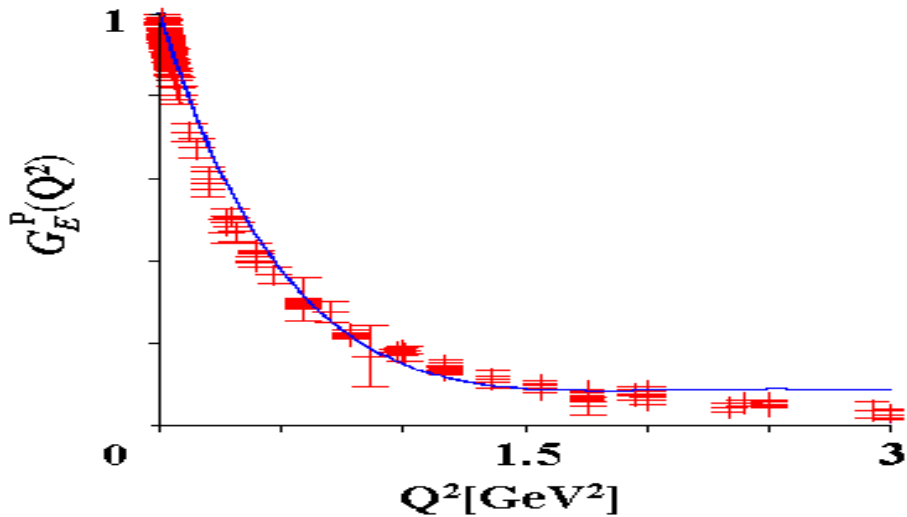}}
\caption{
The proton electric charge form factor in comparison to data. 
The solid line refers to the present calculation in terms of the
solutions  to  eq.~(\ref{chi_ctgnt}). 
The potential parameters are given in the caption of 
Fig.~\ref{levels-Nucleon}. Data compilation same as in ref.~\cite{Berger}. 
\label{Form_F}}
\end{figure}
We obtain the mean square charge radius of the proton-like Roper resonance 
enhanced  by a bit less than 30$\%$  over the proton charge radius. 
Nonetheless, compared to the proton,
the form factor of the  $P_{11}(1440)$ takes smaller
values. This because our predicted $P_{11}(1440)$ charge density
appears slightly arced at origin.
\begin{figure}
\resizebox{0.80\textwidth}{7.5cm}
{\includegraphics{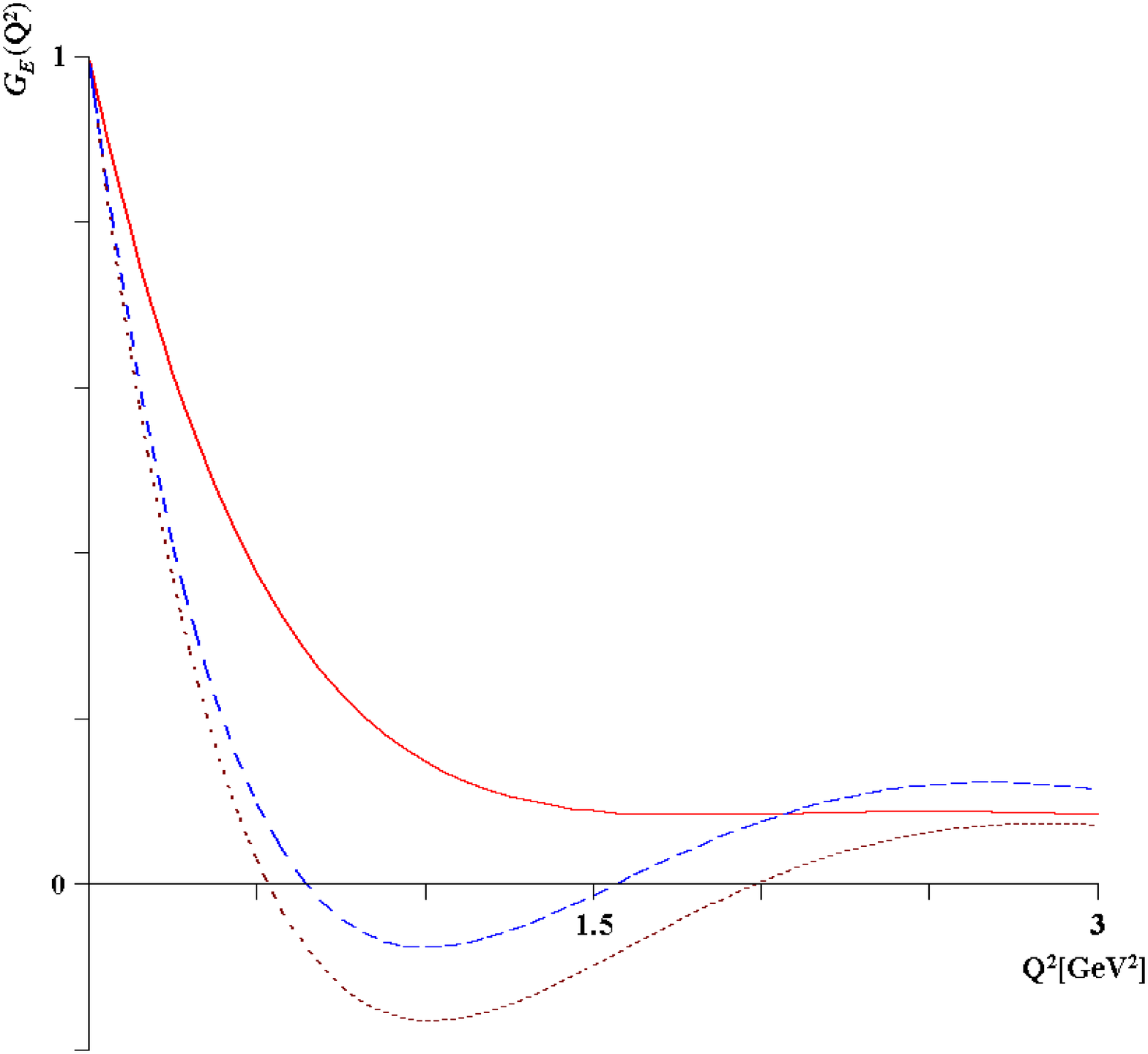}}
\caption{The electric charge form factors of the
 $P_{11}(1440)$ , and $S_{11}(1650)$
resonances (dashed and dotted lines, respectively), in comparison to the
proton electric charge form factor (solid line).  
\label{Formfs}}
\end{figure}

An enhancement, though smaller (10$\%$) has
been found by Nagata and Hosaka in ref.~\cite{Nagata}.
One of the  differences between the model by Nagata-Hosaka 
and the present model is that while in the former  both the 
scalar and axial vector diquarks are weighted by non-zero form factors,
we here  weight them by the extremal 
1 and zero values, respectively. This for the sake of staying as 
close to conformal symmetry as possible. In the present model, it is
the curvature parameter that seems to account for some of the effects
governed by the diquark form factors in flat space quark models.
Curvature as a phenomenological tool suited in simulating   
complicated  many-body effects 
is known to be useful in the description of such complicated many-body
problems as Brownian motion, 
plasma correlations, instanton physics etc. \cite{Brown}, \cite{Nawa}. 
A reason for which  the replacement
of the complicated many-body problem of baryon structure 
(the genuine baryon wave function contains next to 
q-(qq) also 3q, 3q($\bar q$ q)$^n$, 3q(g$^n$) etc configurations) by the
simple q-qq problem on $S^3$ turns out to be a useful approximation to reality
is that curvature, in combination with the conformal gauge 
potential, reasonably accounts for the omitted  many body effects.

In the range of $Q^2\in [0,1.5)$ GeV$^2$, the proton 
form-factor of the present 
relativistic treatment compares in 
quality with $G_E^p(Q^2)$ reported earlier by us in ref.~\cite{CK_07}, where
we employed a $\cot +\csc^2$  confinement potential in
the stationary flat space radial Schr\"odinger equation, though
the relativistic and non-relativistic charge density profiles,
$\sim |\Psi |^2$, 
 in the range of $\chi \in [0, \pi ]$ are quite different 
(see Fig.~\ref{bloed}). 
\begin{figure}
\resizebox{0.80\textwidth}{7.5cm}
{\includegraphics{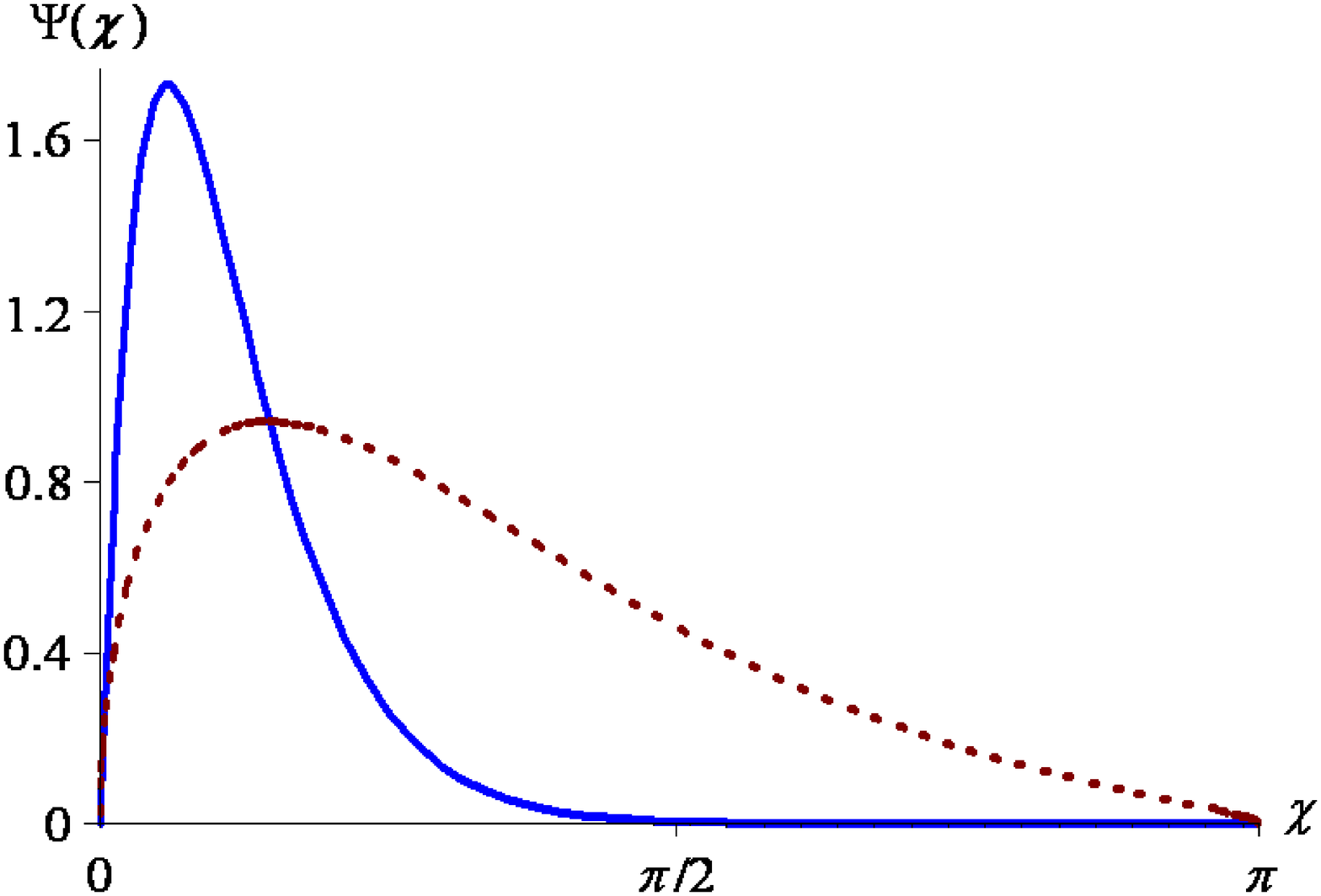}}
\caption{Comparison of the Klein-Gordon (present) and Schr\"odinger 
\cite{CK_07} proton ground state wave functions 
(dashed and solid line, respectively). 
\label{bloed}}
\end{figure}
The profile of the proton charge density
in the non-relativistic 
case is exclusively governed by the cotangent
potential, while in the relativistic case  
it obtains significant contributions from
both the gauge interaction and the gauged centrifugal barrier, 
$\overline{U}_l(\chi, \kappa)$ (defined in 
eqs.~(\ref{new_a}),~(\ref{chi_ctgnt})). The cost for obtaining
nonetheless similar $G_E^p(Q^2)$  form-factors in the above two 
distinct schemes has been
admitting  in the present data fit a larger 
least mean square error in comparison to 
the non-relativistic treatment of ref.~\cite{CK_07}.
Compared  to $G_E^p(Q^2)$  obtained within the 
framework of relativistic quantum mechanics along the line of 
ref.~\cite{Berger} and referred to as GBE CQM there, 
our result on the  mean square proton charge radius appears
somewhat underestimated although  our predicted $G_E^p(Q^2)$  
values fall within the error bars of the reported data below 
2 GeV$^2$, and lie somewhat above afterward. This satisfactory
behavior of the proton electric charge form factor 
is illustrative of the realistic character of the wave functions in 
eq.~(\ref{psi_rad}).

 \section{ Dressing function
for the gluon propagator in the infrared from
Fourier transform of the $S^3$ potential}
This section is devoted to an 
effective instantaneous gluon propagator constructed as
a Fourier transform of the conformal gauge  potential. 
This potential, a cotangent, 
captures quite well the essential traits of QCD dynamics
in so far as it interpolates between the inverse distance potential
(associated with the perturbative regime of
the one gluon exchange) and the infinite well
(associated with asymptotically free though trapped quarks) 
while passing through a region of linear
growth (associated with the non-perturbative regime of flux-tube interactions),
a result due to \cite{CK_07}.
The finite range character of the gauge potential in eq.~(\ref{intr_intr})
is caused by the terms in its
Taylor series decomposition that appear beyond the Coulombic+linear terms
(as seen in eq.~(\ref{crnl})), 
and which can be interpreted as phenomenological non-perturbative corrections
to the former.
For all these reasons, associating the Fourier transform of the cotangent 
function on $S^3$ with an effective gluon propagator seems justified. 
In the following we shall
present this transformation and compare the outcome to 
Lattice QCD results.

\noindent
The gluon, $G_{\mu\nu}^{ab}(q^2)$, and ghost,
$D^{ab}(q^2)$, propagators in the Landau gauge are defined in their turn as
\begin{equation}
G^{ab}_{\mu\nu} =-i
\lbrack 
(
g_{\mu\nu}-
\frac{q_\mu q_\nu }{q^2}
)
\frac{G(q^2)}{q^2} \rbrack\delta^{ab}, 
\label{landau_gauge}
\end{equation}
and
\begin{equation}
D^{ab}(q^2) =\frac{D(q^2)}{q^2}\delta^{ab}.
\end{equation}
Here, $G(q^2)$ and $D(q^2)$ are referred to as the 
respective gluon and ghost dressing functions.
The gluon propagator in this gauge is known to be
transverse in the Lorenz indices, and 
 $\Pi(q)$ stands for the gluon self energy.
Gluon and ghost propagators can be obtained from solving Schwinger-Dyson
equations. In so doing, especially  simple expressions 
for $G(q^2)$ and $D(q^2)$ have been reported in \cite{Fedosenko} as
\begin{equation}
G(q^2)\sim q^2, \quad D(q^2)\sim 1/q^4,
\label{Fed1}
\end{equation}
meaning finiteness of the gluon propagator in the infrared.
Contrary to this, the behavior of the ghost propagator is Coulombic in
the infrared. Both propagators approach zero in the ultraviolet.
More complicated expressions have been calculated in \cite{Aguilar}.
Independently, a finite gluon propagator in the infrared 
has also been  calculated recently 
in lattice QCD \cite{tereza}.

In view of these properties of the gluon propagator, it is of interest to
calculate the gluon dressing function from the {\bf R}$^{1}\otimes S^3$
quark model.  To do so we apply the Born approximation to
$E-V=\sqrt{{\mathbf p}^2 +\mu^2}$
and calculate the instantaneous ($q_0=0$)
gluon propagator in parallel to 
our recent work \cite{CK_09}
as a Fourier transform of the $\cot \chi(r) $ potential on
$S^3$ employing the integration volume in eq.~(\ref{4d_vlm})
in the parametrization of eq.~(\ref{4_PW_FF}).

In Cartesian coordinates the $\cot \chi(r)$ term equals 
$\frac{x_4}{r}$, and stands in fact for two potentials
distinct by a sign and describing interactions on 
the respective Northern, and Southern hemi-spheres.
Correspondingly, their respective  Fourier transforms 
to momentum space  become
\begin{eqnarray}
4\pi \Pi
( |\mathbf{q}| ) &=&-2\mu \frac{(-2G\sqrt{\kappa})}{\hbar ^2} 
\int_0^\infty d|x| |x|^3 \delta (|x| -R)\int_0^{2\pi} d \varphi 
\int_0^\pi 
d\theta \sin\theta \nonumber\\
&&\int_{0/\frac{\pi}{2}}^{\frac{\pi}{2}/\pi } d \chi \sin^2\chi 
e^{i|\mathbf{ q}|\frac{\sin\chi}{\sqrt{\kappa}}|\cos\theta}
\cot \chi ,
\label{b2}
\end{eqnarray}
where the $\delta (|x| -R)$ function restricts {\bf R}$^4$ to $S^3$.
It is the requirement on invertability of this 
transformation, addressed here for the first time, 
that demands for distinguishing between  momentum 
space potentials on the Northern and Southern
hemispheres of $S^3$.
The first potential goes with $\chi \in {\Big[}0,\frac{\pi}{2}{\Big ]}$,  
corresponds to a positive $x_4$, and  describes an increasing 
$r\equiv |{\mathbf r}|\in [0, R]$. The second one
refers
to $\chi \in {\Big[}\frac{\pi}{2},\pi {\Big ]}$, a negative $x_4$, 
and describes a decreasing $|{\mathbf r}|\in [R,0]$. 
To prove the invertability of the integral transform
 it is instructive to cast  the transformation integral
in eq.~(\ref{b2}) to the following equivalent  form,
\begin{eqnarray}
 4\pi \frac{\Pi (|{\mathbf q}|)}{\frac{(4\mu G\sqrt{\kappa}) }
{\hbar ^2} }&=&\pm 
\int_{0}^{R}{\mathrm d}r\frac{r^2}{\sqrt{R^2-r^2}} 
\frac{\sqrt{R^2 -r^2}}{r}
\int_0^{2\pi} d \varphi 
\int_0^\pi 
d\theta \sin\theta e^{i{\mathbf q}\cdot {\mathbf r}}\quad
= \pm 4\pi  \frac{1-\cos |{\mathbf q}| R}{{\mathbf q}^2},\nonumber\\
\label{uw_1}
\end{eqnarray}
where we used the parametrization in eq.~(\ref{4_PW_FF}).
Defining now the integral transform inverse to eq.~(\ref{uw_1}) as,
\begin{eqnarray}
&&\pm \frac{
\sqrt{R^2-r^2}
}{ (2\pi )^3}\int _0^\infty
{\mathrm d}|{\mathbf q}| \, {\mathbf q}^2\int_0^{2\pi }{\mathrm d}\varphi
\int_0^\pi {\mathrm d}\theta \sin \theta  
\frac{4\pi (1-\cos|{\mathbf  q}| R) }{{\mathbf q}^2}
e^{-i{\mathbf q}\cdot {\mathbf  r}}\nonumber\\
&=&\pm\frac{2}{\pi }
\frac{
\sqrt{R^2-r^2}
}{r} \{ 
\begin{array}{cc}
\frac{\pi}{2}, \qquad r<R,\\
\frac{\pi}{4}, \qquad r=R,\\
0, \qquad  r>R,
\end{array}
\end{eqnarray}
proofs the invertability.

We here for concreteness pick up the  Northern 
hemisphere potential  and cast it in the more compact form,
\begin{equation}
\Pi(|{\mathbf q}|)= 
c \frac{2\sin^2 
\frac{|{\mathbf q} |}{2
\hbar \sqrt{\kappa}}}{\left(\frac{|{\mathbf q}|}
{\hbar\sqrt{\kappa }}\right)^2}, \quad c= \frac{4G\mu }{\hbar^2\kappa  }.
\label{prop_we}
\end{equation}
It is increasing  in the infrared,
finite at origin, and approaches the Coulomb 
propagator in the ultraviolet.
In the notations of eq.~(\ref{landau_gauge}) our result 
takes the form
\begin{equation}
\frac{G({\mathbf q} ^2)}{{\mathbf q} ^2}=
c \frac{2\sin^2 
\frac{|{\mathbf q}| }{2}}{{\mathbf q}^2}, 
\end{equation}
for a dimensionless  ${\mathbf q}$  
measured in units of $\hbar\sqrt{\kappa}$. 
Stated differently,
\begin{equation}
G( \mathbf{q} ^2)=2c\sin^2 \frac{|{\mathbf q} |}{2}=
c(1-\cos |{\mathbf q} |)=c
(\frac{{\mathbf q}^2}{2!}-\frac{{\mathbf q}^4}{ 4!}+
\frac{{\mathbf q}^6}{6!}-...),
\end{equation}
and in accord with eq.~(\ref{Fed1}). Therefore,
quark physics in {\bf R}$^{1}\otimes S^3$ also predicts 
a finite gluon dressing function in the infrared which  
approaches zero in the ultraviolet. 
Such a type of behavior has been observed, for example,
in the description of confinement phenomena  \cite{tereza}.
In summary, one of the virtues of the curvature aspect of the 
cotangent gauge potential
is that its $S^3$ Fourier transform comes out well defined.  

\section{Conclusions}

In the present investigation we examined consequences
of conformal symmetry in gravity-gauge duality
on spectroscopic data  on the lightest baryons,
the nucleon and the $\Delta (1232)$. The AdS$_5$/CFT$_4$ concept on
conformal symmetry has been
implemented by a quark-diquark model placed directly
on a conformally
compactified Minkowski spacetime, {\bf R}$^1\otimes S^3$, approached
from the $AdS_5$ cone. 
The description of the $q-(qq)^{0^\pm}$ system on the  
{\bf R}$^1\otimes S^3$ manifold has been executed in terms
of the scalar conformal equation there, gauged by 
a cotangent potential. 
The scalar conformal equation was cast into the form of 
a Klein-Gordon version of the eigenvalue problem of 
the squared  4d angular momentum operator on $S^3$ and presented 
in eq.~(\ref{chi_eq}). 
The spectrum of such a two-body system falls as a whole
into a $\infty$d unitary representation of the conformal group, 
whose levels are irreps of $SO(4)$, the maximal compact group of
$SO(2,4)$. For such $SO(2,4)$ irreps, the notion of  
 ``conformal bands'' has been used.
We observed that all nucleon resonances listed so far by the 
Particle Data Group distribute fairly well over the  first five levels of
a respective conformal band. Same applies to the $\Delta$ resonances.
We identified 38 reported excitations matched by states from the
predicted conformal spectrum and calculated a total of 32 
resonances needed for the completeness of the 
two conformal $N$ and $\Delta$ bands. 
We did not exclude none of  the resonances from the analyzes.
The levels of the conformal band are constituted from 
parity pairs of rising spins of almost equal masses and 
a state of maximal spin of approximately same mass
that remains as a parity simplex.
In this way, more than 54$\%$ of the predicted conformal spectrum has been
matched by  experimentally observed states. 
Finding all the states belonging to the remaining  46$\%$
would provide a compelling  argument
in favor of the realization of conformal symmetry in QCD
in the infrared. However, one should always keep in mind
that the baryonic high-spin states are not stable fundamental particles
but unstable composite  many-body systems,
which can develop a complicated internal dynamics. The latter 
can impose additional conditions on the observability of the
``missing'' states. In that regard suffices to mention threshold and 
cusp effects. Compared to the conformally symmetric
description studied here, the quantum
Hamiltonian of the real resonance systems may
contain higher-order terms of one or more different symmetries. 
In effect,  the irreducible representations of the corresponding
symmetry groups can get mixed up and suppress
some of the quantum numbers. 
Such a situation in physics is by no means new. 
Nuclear physics provides many examples for 
systems in which one part of the spectrum enjoys a symmetry while the 
remaining part either does not, or, has an other symmetry.
Recently, the concept of ``Partial Dynamical Symmetry'' has been 
developed out of the need to address this type of 
peculiarity of many-body excitations \cite{Leviatan}.
Within this new and more relaxed  symmetry context, the 
degeneracy phenomenon in the observed part of the 
baryon spectra investigated here,  already signals relevance
of conformal symmetry in the light flavor baryon spectra 
in line with AdS$_5$/CFT$_{4}$.\\

We also illustrated quality of
the wave functions in calculating realistic values for  charge radii and 
electric-charge form factors of the proton, 
the $P_{11}(1440)$, and $S_{11}(1535)$ states.
We furthermore observed that with the increase of the excitation energies,
when  the influence of the gauge potential gradually fades away,
and the spectrum approaches that of the free conformal rigid rotor, 
the data fit becomes better. Also these observations  point toward 
relevance of conformal symmetry for the spectra of the light flavor baryons.  
Conformal symmetry in the $N$ and $\Delta$ spectra is not an exact symmetry.
The model presented accounts for this circumstance partly through
managing the conformal constant in eq.~(\ref{conf_eq}) 
as a free parameter, and partly through the modification 
of the centrifugal barrier of the
conformally invariant free geodesic motion on $S^3$ through the gauge
interaction. This modification is responsible for the systematic
$P_{2I,1}$--$S_{2I,1}$ mass splitting, which finds a 
satisfactory explanation within the framework under discussion.
Especially the relatively large splittings of about 70 MeV in the 
well established $P_{11}(1440)$--$S_{11}(1535)$, and
$S_{11}(1650)$--$P_{11}(1710)$ pairs have been well reproduced.
As regarding the $S_{11}(2090)$--$P_{11}(2100)$ pair,
we reproduce correctly  the mass ordering but overestimate the splitting.
However, given the poor statistical knowledge on these states
(one star resonances) no conclusion can be drawn from the discrepancy.
The description of conformal symmetry as approximate
is one of the advantages of the Klein-Gordon version (\ref{chi_eq}) 
of the conformal equation on $S^3$
over its Schr\"odinger version in eq.~(\ref{chi_eq_lin})
(earlier considered by us  in ref.~\cite{sprs})
which keeps respecting in the interacting case the degeneracies of the
free geodesic motion.
Encouraging, the reasonable shape of the instantaneous 
effective gluon propagator obtained  as a Fourier transform of 
the cotangent gauge potential. 
To recapitulate, we find  conformal symmetry relevant for 
the spectra of  the lightest baryons.

Finally, a comment on the relevance of the elaborated scheme for mesons is
in order. In ref.~\cite{Afonin_SS} the Crystal Barrel data of the
high-lying non-strange mesons have been analyzed and shown to be 
supportive of the spin-clustering phenomenon suggested in ref.~\cite{MK_97}, 
though the SO(4) levels of the conformal bands
have not been explicitly constructed in ref.~\cite{Afonin_SS}. 
We here fill this gap on the example of
the data below 2100 MeV for purely illustrative purposes. 
The detailed analysis of the meson sector  will be presented elsewhere. 
So far we restrict ourselves to draw the reader's attention 
to Figs.~6 and 7 which depict the population of the
$(3/2,3/2)$ and $(2,2)$ levels of the conformal bands for isoscalar, 
and isovector mesons, respectively. Compared to baryons,
the mesonic $SO(4)$  levels appear parity duplicated which can be read as 
a signal for chiral symmetry restoration from the Goldstone mode
at low energies to the Wigner--Weyl mode at higher energies.  
Therefore, as correctly noticed in ref.~\cite{Afonin_SS}, the Crystal 
Barrel data provide a clear hint on the relevance of both chiral and
conformal symmetry for the lightest flavor mesons at high energies. 
\begin{figure}
\resizebox{0.80\textwidth}{8.01cm}
{\includegraphics{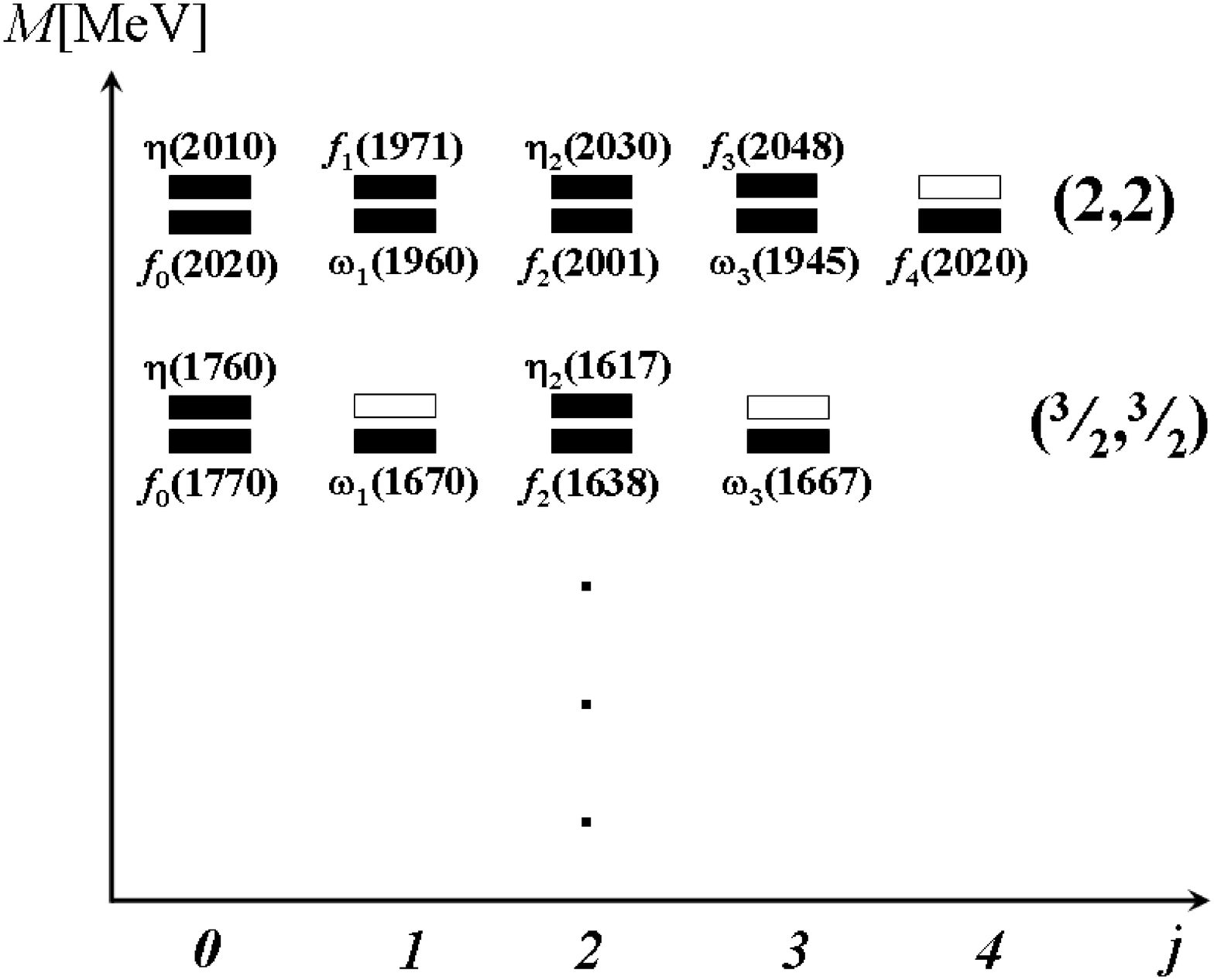}}
\caption{Schematic spectrum of isoscalar non-strange high lying mesons
according to Crystal Barrel data. 
Empty bricks denote ``missing'' states. To the right the
$(K/2, K/2)$ levels from the conformal band with $K=3,4$ 
have been marked.  
The levels appear parity duplicated.
\label{Meson1}}
\end{figure}

\begin{figure}
\resizebox{0.80\textwidth}{8.01cm}
{\includegraphics{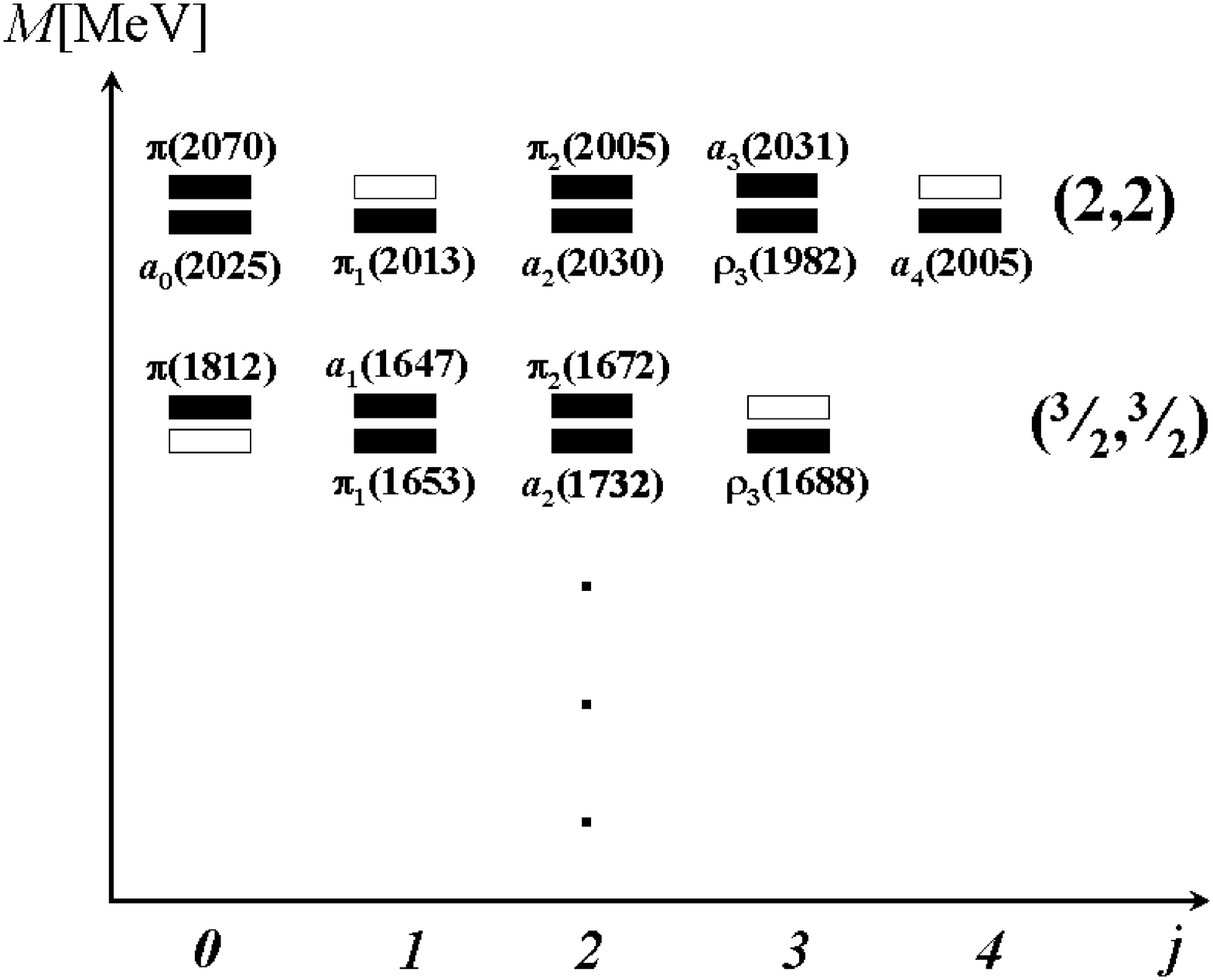}}
\caption{
Schematic spectrum of isovector non-strange high lying mesons according to 
Crystal Barrel data. 
Empty bricks denote ``missing'' states. To the right the
$(K/2, K/2)$ levels from the conformal band with $K=3,4$ have been marked.  
The levels appear parity duplicated.
\label{Meson2}}
\end{figure}

All in all, the model developed in the present work 
provides in our opinion a reasonable 
quantum mechanical approach to QCD  which is congruent  
with the conformal symmetry aspect  of the gravity-gauge
duality.\\

\noindent
Work supported by CONACyT-M\'{e}xico under grant number
CB-2006-01/61286.

\end{document}